\documentclass[preprint,12pt]{elsarticle}

\usepackage{graphicx}

\usepackage{multirow}
\usepackage{subfig}
\usepackage{amsmath}
\usepackage{hyperref}
\biboptions{sort&compress}

\newlength{\dslashwidth}

\newcommand{\beq}{\begin{equation}} 
\newcommand{\eeq}{\end{equation}}
\newcommand{\beqa}{\begin{eqnarray}} 
\newcommand{\eeqa}{\end{eqnarray}}
\newcommand{\newc}{\newcommand}
\newcommand{\bq}{\begin{equation}}
\newcommand{\eq}{\end{equation}}
\newcommand{\ba}{\begin{array}}
\newcommand{\ea}{\end{array}}
\newcommand{\bqa}{\begin{eqnarray}}
\newcommand{\eqa}{\end{eqnarray}}

\newcommand{\lnf}{{\ifmmode \Lambda^{(N_f)} \else $\Lambda^{(N_f)}$\fi}}
\newcommand{\ms}{{\ifmmode \overline{MS} \else $\overline{MS}$\fi}}
\newcommand{\dr}{{\ifmmode \overline{DR} \else $\overline{DR}$\fi}}
\newcommand{\lms}{{\ifmmode \Lambda^{(5)}_{\overline{MS}} \else $\Lambda^{(5)}_{\overline{MS}}$\fi}}
\newcommand{\lam}{{\ifmmode \Lambda \else $\Lambda$\fi}}
\newcommand{\mev}{{\ifmmode {\rm MeV} \else ${\rm MeV}$\fi}}
\newcommand{\gev}{{\ifmmode {\rm GeV} \else ${\rm GeV}$\fi}}
\newcommand{\gevc}{{\ifmmode {\rm GeV/c^2} \else ${\rm GeV/c^2}$\fi}}
\newcommand{\tev}{{\ifmmode {\rm TeV} \else ${\rm TeV}$\fi}}
\newcommand{\tevc}{{\ifmmode {\rm TeV/c^2} \else ${\rm TeV/c^2}$\fi}}
\newcommand{\lp}{{\ifmmode L^+  \else $L^+$\fi}}
\newcommand{\lm}{{\ifmmode L^-  \else $L^-$\fi}}
\newcommand{\mlp}{{\ifmmode M(L^-) \else $M(L^-)$\fi}}
\newcommand{\mlz}{{\ifmmode M(L^0) \else $M(L^0)$\fi}}
\newcommand{\lz}{{\ifmmode L^0 \else $L^0$\fi}}
\newcommand{\ev}{{\ifmmode GeV/c^2 \else $GeV/c^2$\fi}}
\newcommand{\tri}{{\ifmmode \triangleup \else $\triangleup$\fi}}
\newcommand{\unl}{{\ifmmode U_{lL^0} \else $U_{lL^0}$\fi}}\newcommand{\gL}{{\ifmmode g_L \else $g_{L}$\fi}}
\newcommand{\gR}{{\ifmmode g_R  \else $g_{R}$\fi}}
\newcommand{\gumu}{{\ifmmode \gamma^{\mu} \else $\gamma^{\mu}$\fi}}
\newcommand{\gunu}{{\ifmmode \gamma^{\nu} \else $\gamma^{\nu}$\fi}}
\newcommand{\gdmu}{{\ifmmode \gamma_{\mu} \else $\gamma_{\mu}$\fi}}
\newcommand{\gdnu}{{\ifmmode \gamma_{\nu} \else $\gamma_{\nu}$\fi}}
\newcommand{\stw}{{\ifmmode\sin^2\theta_W \else $\sin^{2}\theta_{W}$ \fi}}
\newcommand{\sws}{{\ifmmode \;\sin^2\theta_W  \else $\;\sin^{2}\theta_{W}$ \fi}}
\newcommand{\cws}{{\ifmmode \;\cos^2\theta_W  \else $\;\cos^{2}\theta_{W}$ \fi}}
\newcommand{\sw}{{\ifmmode \;\sin\theta_W  \else $\sin\theta_{W}$ \fi}}
\newcommand{\cw}{{\ifmmode \;\cos\theta_W  \else $\;\cos\theta_{W}$ \fi}}
\newcommand{\tw}{{\ifmmode \;\tan\theta_W  \else $\;\tan\theta_{W}$ \fi}}
\newcommand{\qq}{{\ifmmode q\overline{q} \else $q\overline{q}$\fi}}
\newcommand{\lR}{{\ifmmode l_R \else $l_R$\fi}}
\newcommand{\lL}{{\ifmmode l_L \else $l_L$\fi}}
\newcommand{\nt}{{\ifmmode \nu_{\tau} \else $\nu_{\tau}$\fi}}
\newcommand{\nuR}{{\ifmmode \nu_R  \else $\nu_R$\fi}}
\newcommand{\nuL}{{\ifmmode \nu_L  \else $\nu_L$\fi}}
\newcommand{\qR}{{\ifmmode g_R  \else $q_R$\fi}}
\newcommand{\qL}{{\ifmmode q_L  \else $q_L$\fi}}
\newcommand{\qRp}{{\ifmmode q_R'  \else $q_{R}$'\fi}}
\newcommand{\qLp}{{\ifmmode q_L'  \else $q_{L}$'\fi}}
\newcommand{\est}{{\ifmmode e^{\bf \ast} \else $e^{\bf \ast}$\fi}}
\newcommand{\lst}{{\ifmmode l^{\bf \ast} \else $l^{\bf \ast}$\fi}}
\newcommand{\must}{{\ifmmode \mu^{\bf \ast} \else $\mu^{\bf \ast}$\fi}}
\newcommand{\taust}{{\ifmmode \tau^{\bf \ast} \else $\tau^{\bf \ast}$ \fi}}
\newcommand{\pperp}{{\ifmmode p_t  \else $p_t$\fi}}
\newcommand{\et}{{\ifmmode E_t  \else $E_t$\fi}}
\newcommand{\xt}{{\ifmmode x_t  \else $x_t$\fi}}
\newcommand{\smumu}{{\ifmmode \sigma_{\mu\mu}  \else $\sigma_{\mu\mu}$ \fi}}
\newcommand{\eg}{{\ifmmode e\gamma  \else $e\gamma$\fi}}
\newcommand{\epem}{{\ifmmode e^+e^-  \else $e^+e^-$\fi}}
\newcommand{\lplm}{{\ifmmode L^+L^-  \else $L^+L^-$\fi}}
\newcommand{\pp}{{\ifmmode p\overline p  \else $p\overline p$\fi}}
\newcommand{\llz}{{\ifmmode L^0\overline{L}^0 \else $L^0\overline{L}^0$\fi}}
\newcommand{\epemt}{{\ifmmode e^+e^- \to  \else $e^+e^- \to$\fi}}
\newcommand{\eb}{{\ifmmode E_{beam}  \else $E_{beam}$\fi}}
\newcommand{\ip}{{\ifmmode pb^{-1}  \else $pb^{-1}$\fi}}
\newcommand{\upm}{{\ifmmode ^{\pm}  \else $^{\pm}$\fi}}
\newcommand{\de}{{\ifmmode ^{\circ}  \else $^{\circ}$ \fi}}
\newcommand{\appr}{{\ifmmode \sim \else $\sim$ \fi}}
\newcommand{\corresp}{{\ifmmode \stackrel{\wedge}{=} \else $\stackrel{\wedge}{=}$ \fi}}
\newcommand{\sqrts}{{\ifmmode \sqrt{s} \else $\sqrt{s}$\fi}}
\newcommand{\zz}{{\ifmmode Z^0  \else $Z^0$\fi}}
\newcommand{\mz}{{\ifmmode M_{Z}  \else $M_{Z}$\fi}}
\newcommand{\mzs}{{\ifmmode M_{Z}^2  \else $M_{Z}^2$\fi}}
\newcommand{\mw}{{\ifmmode M_{W}  \else $M_{W}$\fi}}
\newcommand{\mws}{{\ifmmode M_{W}^2  \else $M_{W}^2$\fi}}
\newcommand{\mh}{{\ifmmode M_{Higgs}  \else $M_{Higgs}$\fi}}
\newcommand{\msusy}{{\ifmmode M_{SUSY}  \else $M_{SUSY}$\fi}}
\newcommand{\msusys}{{\ifmmode M_{SUSY}^2  \else $M_{SUSY}^2$\fi}}
\newcommand{\su}{{\ifmmode SU(3)_C\otimes\- SU(2)_L\otimes\- U(1)_Y \else $SU(3)_C\otimes\A0SU(2)_L\otimes U(1)_Y$\fi}}
\newcommand{\suthree}{{\ifmmode SU(3)_C  \else $SU(3)_C$\fi}}
\newcommand{\sutwo}{{\ifmmode  SU(2)_L\otimes U(1)_Y \else $SU(2)_L\otimes U(1)_Y$\fi}}
\newcommand{\taup}{{\ifmmode \tau_{proton} \else $\tau_{proton}$\fi}}
\newcommand{\as}{{\ifmmode \alpha_{s}  \else $\alpha_{s}$\fi}}
\newcommand{\mgut}{{\ifmmode M_{GUT}  \else $M_{GUT}$\fi}}
\newcommand{\mguts}{{\ifmmode M_{GUT}^2  \else $M_{GUT}^2$\fi}}
\newcommand{\mzero}{{\ifmmode m_0        \else $m_0$\fi}}
\newcommand{\mhalf}{{\ifmmode m_{1/2}    \else $m_{1/2}$\fi}}
\newcommand{\sq}{{\ifmmode \tilde{q}    \else $\tilde{q}$\fi}}
\newcommand{\gl}{{\ifmmode \tilde{g}    \else $\tilde{g}$\fi}}
\newcommand{\mb}{{\ifmmode m_{b}    \else $m_{b}$\fi}}
\newcommand{\mt}{{\ifmmode m_{t}    \else $m_{t}$\fi}}
\newcommand{\mts}{{\ifmmode m_{t}^2    \else $m_{t}^2$\fi}}

\newcommand{\mtau}{{\ifmmode m_{\tau}  \else $m_{\tau}$\fi}}
\newcommand{\dpp}{{\ifmmode \delta_{pert} \else $\delta_{pert}$\fi}}
\newcommand{\dnp}{{\ifmmode\delta_{non-pert}\else$\delta_{non-pert}$\fi}}
\newcommand{\dew}{{\ifmmode \delta_{\rm EW}\else $\delta_{\rm EW}$\fi}}
\newcommand{\rt}{{\ifmmode R_{\tau}  \else $R_{\tau} $\fi}}
\newcommand{\rz}{{\ifmmode R_{Z}  \else $R_{Z} $\fi}}

\newcommand{\swb}{{\ifmmode \sin^2\theta_{\overline{MS}} \else $\sin^2\theta_{\overline{MS}}$\fi}}
\newcommand{\cwb}{{\ifmmode \cos^2\theta_{\overline{MS}} \else $\cos^2\theta_{\overline{MS}}$\fi}}

\newc\AIPCP[3] {{\em AIP Conf. Proc.} {\bf #1} (#2) #3}
\newc\AJ[3] {{\em Astrophys. J.} {\bf #1} (#2) #3}
\newc\AMS[3] {{\em Ann. Math. Statist.} {\bf #1} (#2) #3}
\newc\AP[3] {{\em Ann. Phys.} {\bf #1} (#2) #3}
\newc\APJ[3] {{\em Astropart. J.} {\bf #1} (#2) #3}
\newc\APP[3] {{\em Astropart. Phys.} {\bf #1} (#2) #3}
\newc\APS[3] {{\em Astrophys. J. Suppl.} {\bf #1} (#2) #3}
\newc\ARNPS[3] {{\em Ann. Rev. Nucl. Part. Sci.} {\bf C#1} (#2) #3}
\newc\BA[3] {{\em Bayesian Anal.} {\bf C#1} (#2) #3}
\newc\CPC[3] {{\em Comput. Phys. Commun.} {\bf C#1} (#2) #3}
\newc\CP[3] {{\em Contemp. Phys.} {\bf #1} (#2) #3}
\newc\EPJ[3] {{\em Euro. Phys. Journ.} {\bf C#1} (#2) #3}
\newc\JCAP[3] {{\em JCAP} {\bf #1} (#2) #3}
\newc\JHEP[3] {{\em JHEP} {\bf #1} (#2) #3}
\newc\JPG[3] {{\em J. Phys.} {\bf G #1} (#2) #3}
\newc\IJMP[3] {{\em Int. J. Mod. Phys.} {\bf A #1} (#2) #3}
\newc\MNRAS[3] {{\em Mon. Not. Roy. Astron. Soc.} {\bf #1} (#2) #3}
\newc\MPL[3] {{\em Mod. Phys. Lett.} {\bf A #1} (#2) #3}
\newc\NAR[3] {{\em New Astron. Rev.} {\bf #1} (#2) #3}
\newc\NCA[3] {{\em Nuovo Cimento} {\bf #1} (#2) #3}
\newc\NIM[3] {{\em Nucl. Instrum. Methods} {\bf #1} (#2) #3}
\newc\NIMA[3] {{\em Nucl. Instrum. Methods} {\bf A #1} (#2) #3}
\newc\NAT[3] {{\em Nature} {\bf #1} (#2) #3}
\newc\NPB[3] {{\em Nucl. Phys.} {\bf B #1} (#2) #3}
\newc\NPA[3] {{\em Nucl. Phys.} {\bf A #1} (#2) #3}
\newc\NPPS[3] {{\em Nucl. Phys. Proc. Suppl.} {\bf #1} (#2) #3}
\newc\PLB[3] {{\em Phys. Lett.} {\bf B #1} (#2) #3}
\newc\PR[3] {{\em Phys. Rep.} {\bf #1} (#2) #3}
\newc\PRL[3] {{\em Phys. Rev. Lett.} {\bf #1} (#2) #3}
\newc\PRD[3] {{\em Phys. Rev.} {\bf D #1} (#2) #3}
\newc\PRC[3] {{\em Phys. Rev.} {\bf C #1} (#2) #3}
\newc\PTP[3] {{\em Prog. Theor. Phys.} {\bf #1} (#2) #3}
\newc\RMP[3] {{\em Rev. Mod. Phys.} {\bf #1} (#2) #3 }
\newc\RPP[3] {{\em Rept. Prog. Phys.} {\bf #1} (#2) #3 }
\newc\SC[3] {{\em Science} {\bf #1} (#2) #3 }
\newc\ZPC[3] {{\em Z. Phys.} {\bf C #1} (#2) #3}
\newc\Err[3] {{\em Erratum-ibid.} {\bf #1} (#2) #3 }

\journal{Physics Letters B}

\begin{document}

\begin{frontmatter}


%
\title{Can we discover a light singlet-like NMSSM Higgs boson at the LHC?}

\author[label1]{C. Beskidt}\ead{conny.beskidt@kit.edu}
\author[label1]{W. de Boer}\ead{wim.de.boer@kit.edu}
\author[label1,label2]{D.I. Kazakov}\ead{KazakovD@theor.jinr.ru} 
\address[label1]{Institut f\"ur Experimentelle Teilchenphysik, Karlsruhe Institute of Technology, P.O. Box 6980, 76128 Karlsruhe, Germany}
\address[label2]{Bogoliubov Laboratory of Theoretical Physics, Joint Institute for Nuclear Research, 141980, 6 Joliot-Curie, Dubna, Moscow Region, Russia}

\begin{abstract}

In the next-to minimal supersymmetric standard model (NMSSM) one additional singlet-like Higgs boson with small couplings to standard model (SM) particles is introduced. Although the mass can be well below the discovered 125 GeV Higgs boson mass its small couplings may make a discovery at the LHC difficult.
We use a novel scanning technique to efficiently scan the whole parameter space and determine the range of cross sections and branching ratios for the light singlet-like Higgs boson below 125 GeV.
This allows to determine the perspectives for the future discovery potential at the LHC. Specific LHC benchmark points are selected representing the salient NMSSM features.

\end{abstract}

\end{frontmatter}


\section{Introduction}
\label{Introduction}

Supersymmetry (SUSY) predicts a light Higgs boson with a mass below 130 GeV (for reviews see \cite{Haber:1984rc,deBoer:1994dg,Martin:1997ns}) which is compatible with the discovered Higgs-like boson with SM-like couplings and a mass of 125 GeV \cite{Aad:2012tfa,Chatrchyan:2012ufa}. In addition to the SM-like Higgs boson a second singlet-like Higgs boson is predicted in the next-to-minimal supersymmetric standard model (NMSSM) \cite{Ellwanger:2009dp}. This additional Higgs boson couples only weakly to SM particles because of its large singlet content. So the decay modes for the singlet-like Higgs boson differ from the well-known decays of the SM Higgs boson. In addition, the singlet-like couplings lead to a small production cross section. 

The introduction of an additional Higgs singlet $S$ in the NMSSM yields more parameters in the Higgs sector for the interactions between the singlet and the Higgs doublets and the singlet self interaction. 
Even if one considers the well-motivated subspace with unified masses and couplings at the GUT scale the additional particles and their interactions lead to a large parameter space. To cope with this large parameter space and especially the large correlations between the parameters, we use a novel scanning technique to obtain the expected range of cross sections and branching ratios of the light singlet-like Higgs boson. This method was previously used for the heavy Higgs boson \cite{Beskidt:2016egy} and will be shortly described in Sect. \ref{analysis}. In this letter we apply this method, which allows for an efficient scanning of the whole parameter space with a complete coverage, to the light singlet-like Higgs boson and determine the cross sections and branching ratios over the whole parameter space, thus complementing previous studies using methods not guaranteeing complete coverage \cite{Guchait:2015owa,King:2014xwa,Cao:2016uwt,King:2012is,Potter:2015wsa,Bomark:2015hia,Cao:2013gba,Badziak:2013bda,Ellwanger:2012ke,Gunion:2012zd,
Dermisek:2008uu,Bernon:2014nxa,Muhlleitner:2017dkd,Das:2016eob,Barbieri:2013nka,Mariotti:2017vtv,Baum:2017gbj,Bandyopadhyay:2015tva}.
The singlet-like Higgs boson can be the lightest Higgs boson $H_1$ implying it has a mass below 125 GeV, although scenarios, where the SM-like Higgs boson is the lightest one, are also possible. However, since a singlet-like Higgs boson has by definition small couplings to SM-like particles we concentrate on $m_{H1}  < 125$ GeV, where the phase-space and correspondingly, the cross section can still be large despite the small couplings. 
An interesting possibility is the fact that the slight excess of a Higgs-like signal seen at 98 GeV at the LEP originates from the $H_1$
Higgs boson as discussed in Ref. \cite{Belanger:2012tt} after the Higgs boson discovery or even before \cite{Dermisek:2007yt}.
After a short summary of the Higgs sector in the NMSSM we summarize the fit strategy to sample the NMSSM parameter space based on the 3D neutral Higgs boson mass space. We find two regions for the couplings of the singlet-like Higgs boson to itself (called $\kappa$) and to the other Higgs bosons (called $\lambda)$, namely regions with large (small) values of $\lambda$ and $\kappa$, which are called Region I (II), respectively.
The Higgs singlet production has been studied before in Ref. \cite{Ellwanger:2015uaz} as well using also the distinction between these two regions with the focus on $\gamma\gamma$ final state. With our novel scanning technique yielding complete coverage we can study in detail the branching ratios of all channels and discover large differences between the two Regions.
We conclude by showing the branching ratios and cross sections times branching ratios as function of the Higgs boson mass for the most promising discovery channels like $\tau\tau, \gamma\gamma, Z\gamma, ZZ, WW, \tilde{\chi}_0^1\tilde{\chi}_0^1$ and $A_1 A_1$.
We select benchmark points in 4 bins of the Higgs boson mass $m_{H1}$ in both, Regions I and II, for each of the most promising discovery channels. These benchmark points, as detailed in the Appendix, can be used to simulate the discovery channels and its background more precisely in order to get a quantitative determination of the discovery potential.

\section{NMSSM Higgs sector}
\label{higgs}
We focus on the well-motivated  semi-constrained NMSSM, as described in Ref. \cite{Ellwanger:2009dp} and  use the corresponding code NMSSMTools 5.2.0 \cite{Das:2011dg} to calculate the SUSY mass spectrum, Higgs boson masses and  branching ratios from the NMSSM parameters.  The Higgs production cross sections are calculated with SusHi \cite{Harlander:2012pb,Harlander:2002wh,Harlander:2003ai,Aglietti:2004nj,Bonciani:2010ms,Degrassi:2010eu,Degrassi:2011vq,Degrassi:2012vt,Liebler:2015bka}.

 Within the NMSSM the Higgs fields consist of the two Higgs doublets  ($H_u, H_d$), which appear in the MSSM as well, but in addition, the NMSSM has an additional complex Higgs singlet $S$. 
Furthermore, we have the GUT scale parameters of the constrained minimal supersymmetric standard model (CMSSM): $\mzero$, $\mhalf$ and $A_0$, where $\mzero$($\mhalf$) are the common mass scales of the spin 0(1/2) SUSY particles at the GUT scale and $A_0$ is the trilinear coupling of the CMSSM Higgs sector at the GUT scale. In total, the semi-constrained NMSSM has nine free parameters:
\begin{equation}
 \mzero,~ \mhalf,~A_0,~ \tan\beta,  ~ \lambda, ~\kappa,  ~A_\lambda, ~A_\kappa, ~\mu_{eff}.
\label{params}
\end{equation}
Here $\tan\beta$ corresponds to the ratio of the vevs of the Higgs doublets, i.e. $\tan\beta\equiv v_u/v_d$,
$ \lambda$ represents the coupling between the Higgs singlet and doublets ($\lambda S H_u\cdot  H_d$),  $~\kappa$ the self-coupling of the singlet ($\kappa S^3/3$);   $A_\lambda$ and $A_\kappa$ are the corresponding trilinear soft breaking terms, $\mu_{eff}$ represents an effective Higgs mixing parameter and is related to the vev of the singlet $s$ via the coupling $\lambda$, i.e. $\mu_{eff}\equiv\lambda s$. Therefore, $\mu_{eff}$ is naturally of the order of the electroweak scale \cite{King:2012tr,Cao:2012fz}, thus avoiding the $\mu$-problem \cite{Ellwanger:2009dp}. The latter six parameters in Eq. \ref{params} form the 6D parameter space of the NMSSM Higgs sector. $A_0$ is highly correlated with $A_\lambda$ and $A_\kappa$ in the semi-constrained NMSSM, so fixing it would restrict the range of $A_\lambda$ and $A_\kappa$ severely. Therefore, $A_0$ is allowed to vary as well, which leads in total to 7 free parameters and thus a 7D parameter space.

The neutral components from the two Higgs doublets and singlet  mix to form three physical CP-even
 scalar bosons and two physical CP-odd pseudo-scalar bosons.
The elements of the corresponding mass matrices at
tree level  are given in Ref. \cite{Miller:2003ay}.
The mass eigenstates of the neutral Higgs bosons are determined by the diagonalization of the mass matrix, so the scalar Higgs bosons $H_i$, where the index $i$ increases with increasing mass, are mixtures of the CP-even weak eigenstates $H_d, H_u$ and $S$ 
\begin{eqnarray}\label{eq1}
H_i=S_{i1}  H_d  + S_{i2}  H_u  + S_{i3}  S, 
\end{eqnarray}
where $S_{ij}$ with $i,j=1,2,3$ are the elements of the Higgs mixing matrix.
For the lightest Higgs boson with $i=1$ the value of $S_{13}$ is usually close to 1, which implies small couplings of $H_1$ to SM particles as will be discussed below.
The Higgs couplings to quarks and leptons of the third generation are crucial for the allowed range of branching ratios and given by:
\begin{align}
H_i t_L t_R^c &:  -\frac{h_t}{\sqrt{2}}S_{i2} & h_t &= \frac{m_t}{v \sin\beta},\notag\\
H_i b_L b_R^c &: \frac{h_b}{\sqrt{2}}S_{i1} & h_b &= \frac{m_b}{v \cos\beta},\label{coupling}\\
H_i \tau_L \tau_R^c &: \frac{h_\tau}{\sqrt{2}}S_{i1} & h_\tau &= \frac{m_\tau}{v \cos\beta},\notag
\end{align}
where $h_t$, $h_b$ and $h_\tau$ are the corresponding Yukawa couplings. The relation includes the quark and lepton masses $m_t$, $m_b$ and $m_\tau$ and $v^2=v_u^2+v_d^2$. The couplings to fermions of the first and second generation are analogous to Eq. \ref{coupling} with different quark and lepton masses. 

\begin{figure}
\begin{center}
\hspace{-1cm}
\includegraphics[width=0.98\textwidth]{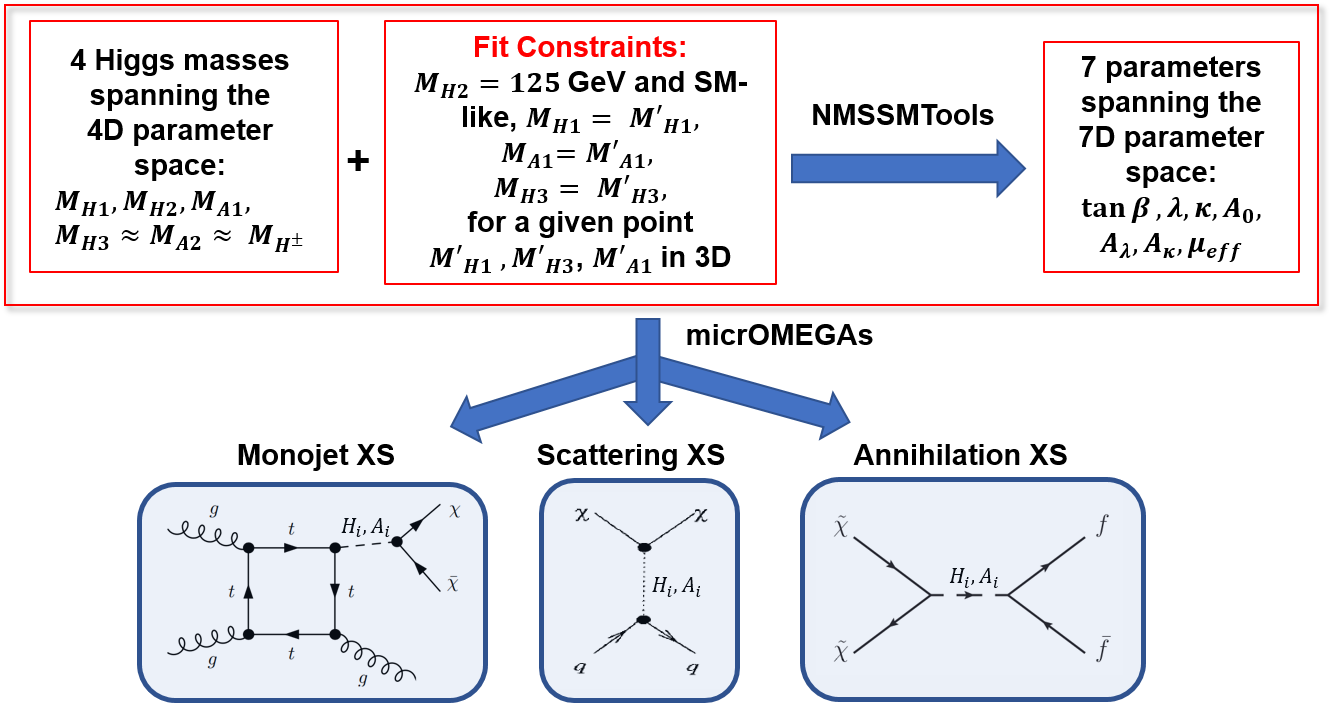}
\caption[]{ 
Schematic diagram of the scanning technique to determine the branching ratios as function of the Higgs boson masses. Scanning strategy for accepted points: select 4 Higgs masses (left box), fit these masses with the 7 free NMSSM parameters (right box) using the following constraints: $M_{H2}$ = 125 GeV with SM-like couplings to quarks, leptons and gauge bosons (9 constraints), apply LHC and LEP Higgs mass limits. The relation between the NMSSM parameters and masses is encoded in NMSSMTools. Repeat the fit in a grid of all Higgs mass combinations of $M_{A1}, M_{H1}, M_{H3}$ (for $M_{A1} <$ 500 GeV and $M_{H3} <$ 2 TeV) to obtain a scan over all accepted NMSSM parameters in the 7D parameter space. From the 7 parameters the dark matter cross sections can be calculated with micrOMEGAs for each accepted point. 
Scanning the parameter space with full coverage means scanning over a grid of $M_{H1}$, $M_{H3}$ and $M_{A1}$ masses and performing for each mass combination the fit to determine the NMSSM parameters. 
Studying the influence of e.g the $M_{H3}$ mass by marginalizing over the $M_{H1}$ and $M_{A1}$ mass can be done by scanning in a plane with a constant $M_{H3}$ mass and repeating this for different values of $M_{H3}$. 
}
\label{method}
\end{center}
\end{figure}

\section{Analysis}
\label{analysis}

The branching ratios and cross sections of the light Higgs boson have been determined for two different regions, since a certain Higgs mass combination is not unique, as can be easily seen already from the approximate expression for the 125 GeV Higgs boson \cite{Ellwanger:2009dp}: 
\beq\label{eq4}
M_{H}^2\approx M_Z^2\cos^2 2\beta+ \Delta_{\tilde{t}} + \lambda^2 v^2 \sin^2 2\beta - \frac{\lambda^2}{\kappa^2}(\lambda-\kappa \sin 2 \beta)^2.  
\eeq 
The first tree level term can become at most $M_Z^2$ for large $\tan\beta$. The diffe\-rence between $M_Z$ and 125 GeV has to originate mainly from the lo\-garithmic stop mass corrections $\Delta_{\tilde{t}}$. The two remaining terms originate from the mixing with the singlet of the NMSSM at tree level and become large for large values of the couplings $\lambda$ and $\kappa$ and small $\tan\beta$. As mentioned before, this region we call \textit{Region I}. However, there exists another solution to Eq. \ref{eq4} with small values of $\lambda,\kappa$ and large values of $\tan\beta$. 
This we call \textit{Region II} (also mentioned before), which can be obtained by a trade-off between the first two terms and last two terms. 
So Region II with its small couplings $\lambda$ and $\kappa$ is in some sense closer to the MSSM although the singlet-like Higgs and its corresponding singlino-like LSP yield additional physics, like the possibility of double Higgs production and an LSP hardly coupling to matter.
In both regions the radiative corrections from stop loops can be small with stop masses around the TeV scale. Quantitatively, Region I is defined by $\lambda > 0.3, \tan\beta < 10$  and Region II by $\lambda < 0.1, \tan\beta < 30$. These upper and lower limits for $\lambda$ and $\tan\beta$ were suggested by the $\chi^2$ distribution of Fig. 1 in Ref. \cite{Beskidt:2016egy}. The limit for $\tan\beta$ in Region II allows additionally to be consistent with the results from B-physics.

For each set of the 7 parameters in the Higgs sector the 6 Higgs boson masses are completely determined:  3 scalar Higgs masses $m_{H_i}$, 2 pseudo-scalar Higgs masses $m_{A_i}$ and the charged Higgs boson mass $m_{H^\pm}$. The masses of $A_2$, $H_3$ and $H^\pm$ are of the order of $M_A$, if $M_A >> M_Z$. Then only one of the masses  is needed. Furthermore, either $H_1$ or $H_2$ has to be the observed Higgs boson with a mass of 125 GeV, so there are only 3 free neutral Higgs boson masses in the NMSSM, i.e. a 3D parameter space, e.g. $m_{A_1}$, $m_{H_1}$ and $m_{H_3}\approx m_{A_2} \approx m_{H^\pm}$. We choose $m_{H_2}=125$ GeV, so $m_{H_1}<125$ GeV. 
Instead of scanning over the 7D parameter space of the NMSSM parameters to determine the range of Higgs boson masses, as was done by other groups in the (N)MSSM, see e.g. Ref. \cite{Arganda:2012qp,Arganda:2013ve,Muhlleitner:2017dkd}, one can invert the problem and scan the 3D parameter space of the Higgs boson masses. 
For each combination of Higgs boson masses one finds a single set in the 7D parameter space of the NMSSM parameters. This is graphically illustrated in Fig. \ref{method}. The transition of the 3D to 7D parameter space can be done by a Minuit \cite{James:1975dr} fit with the constraints given in the upper middle box of Fig. \ref{method}. The connection between the upper left and right box is obtained from NMSSMTools 5.2.0.. Note that the fit is free to determine the optimum values of the parameters from the top right box in Fig. \ref{method} within the corresponding range of Regions I and II for each combination of Higgs boson masses. 
The $\chi^2$ function to be minimized includes the following contributions 
\beq\label{eq5}
\chi^2_{tot}=\chi^2_{H_1}+\chi^2_{H_2}+\chi^2_{H_3}+\chi^2_{A_1}+\chi^2_{LEP}+\chi^2_{LHC}(+\chi^2_{\Omega h^2}+\chi^2_{B-physics}+\chi^2_{DDMS}).
\eeq 
The terms $\chi^2_{A_1}$ and $\chi^2_{H_i}$ for $i=1,3$ require the NMSSM parameters to be adjusted such that the masses of the Higgs bosons $m_{H_{1/3}}$ and $m_{A_1}$ agree with the chosen point in the 3D mass space. The error $\sigma_{A_1/H_i}$ is set to 2 GeV. This leads to a large fluctuation of the mass points for light Higgs boson masses below 10 GeV. However, it was checked that this does not impact the results of the branching ratio range. Since the lightest Higgs boson $H_1$ has a mass below 125 GeV, the LEP constraints on the couplings of a light Higgs boson below 114 GeV, as obtained from Higgs searches at LEP (Fig. 10 in Ref. \cite{Schael:2006cr}), are included. Additionally, the LEP limit on the chargino mass is applied and both constraints are represented by $\chi^2_{LEP}$, as listed in Ref. \cite{Beskidt:2014kon}. 
These constraints are in principle implemented in NMSSMTools, but small corrections were applied. The second lightest Higgs boson corresponds to the observed Higgs boson with couplings close to the SM couplings. These constraints are included in the term  $\chi^2_{H_2}$ which implies 8 additional constraints by requiring the couplings to quarks, leptons and gauge bosons to be compatible with the standard model couplings. This was implemented by requiring the 8 scaling parameters for the corresponding cross sections in NMSSMTools to become 1. More details about the $\chi^2$ contributions have been spelled out in Ref. \cite{Beskidt:2016egy}. In addition, constraints from the LHC as implemented in NSSMTools concerning light scalar and pseudo-scalar Higgs bosons, see Refs. \cite{Khachatryan:2015nba,Aad:2015oqa,Khachatryan:2015wka}, are included as well represented by $\chi^2_{LHC}$. The analysis can be easily expanded by additional constraints from the dark matter sector like the relic density $\Omega h^2$ and direct dark matter searches (DDMS) (bottom boxes in Fig. \ref{method}) calculated with micrOMEGAs \cite{Belanger:2013oya} and/or B-physics results. The corresponding $\chi^2$ contributions can be added, as indicated by the terms in brackets in Eq. \ref{eq5}.  
We define the range of the Higgs masses in the 3D mass space as follows:
\begin{align}\label{range}
5~GeV <  & ~m_{H_1} < 125~GeV, \notag\\
125~GeV <  & ~m_{H_3} < 2~TeV, \\
5~GeV <  & ~m_{A_1} < 500~GeV.  \notag
\end{align}
Fitting for all selected Higgs mass combinations yields the optimal couplings, and hence the Higgs mixing matrix for each Higgs mass combination,  which determines the branching ratios of all 6 Higgs bosons.  

The advantage of scanning the 3D mass space instead of the 7D para\-meter space can be appreciated as follows:
Instead of systematically scanning the whole NMSSM parameter space one usually resorts to a reduced set of para\-meter combinations using random scans, but one never can be sure about the coverage because of the high correlations between the parameters. A highly correlated parameter space cannot be efficiently sampled without taking a correlation matrix into account. The correlation matrix tells how to step through the parameter space in a correlated way but the correlations are not known and difficult to determine in a multi-dimensional space. 
By scanning the 3D mass space in the chosen range of largely uncorrelated Higgs masses (Eq. \ref{range}) the corresponding parameter space of the couplings is covered. We compared with results using the general NMSSM (see e.g. Ref. \cite{Muhlleitner:2017dkd}) and find similar branching ratios and cross sections, but obtain more insight by separating the acceptable coup\-lings in Regions I and II, as will be discussed in Sect. \ref{br-higgs}. Surprisingly, the NMSSM-like Region I has branching ratios mainly to b-quarks and tau-leptons, as expected in the MSSM, while
the MSSM-like Region II has regions with zero $H_1$-couplings to b-quarks. The different behaviour of the two Regions can be
understood by reconstructing the Higgs mixing matrix from the fitted couplings using a method with full coverage of the parameter
space.

It should be noted that the results of this letter are not sensitive to the restriction to the constrained NMSSM since the Higgs sector is mostly independent of $m_0$ and $m_{1/2}$, which enter only in the stop corrections $\Delta_{\tilde{t}}$ in Eq. \ref{eq4}. 
Since the Higgs mass dependence on the stop mass is logarithmic, a different stop mass leads to small shifts in the optimal values of the NMSSM parameters in the upper right panel of Fig. \ref{method}. 
This was checked by changing the values of $m_0$ and $m_{1/2}$. However, choosing the constrained model reduces the number of free parameters and additionally, allows to use the full radiative corrections from the unification scale to the weak scale to all masses and couplings and introduce electroweak symmetry breaking. For this reason, the values of $m_0$ and $m_{1/2}$ have been fixed to 1 TeV, which is consistent with the current LHC limits \cite{Aad:2015iea}.

\begin{figure}
\begin{center}
\hspace{-1cm}
\includegraphics[width=0.48\textwidth]{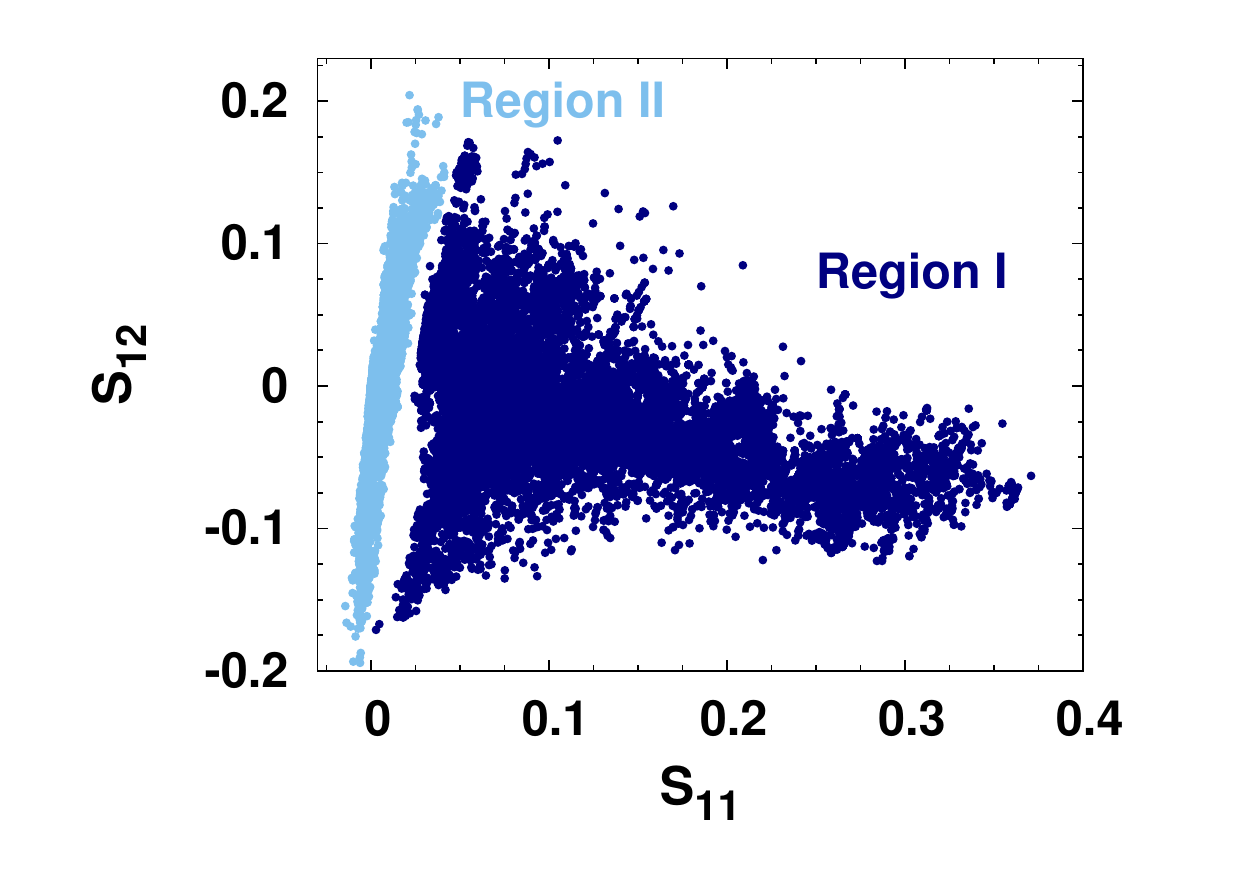}
\includegraphics[width=0.55\textwidth]{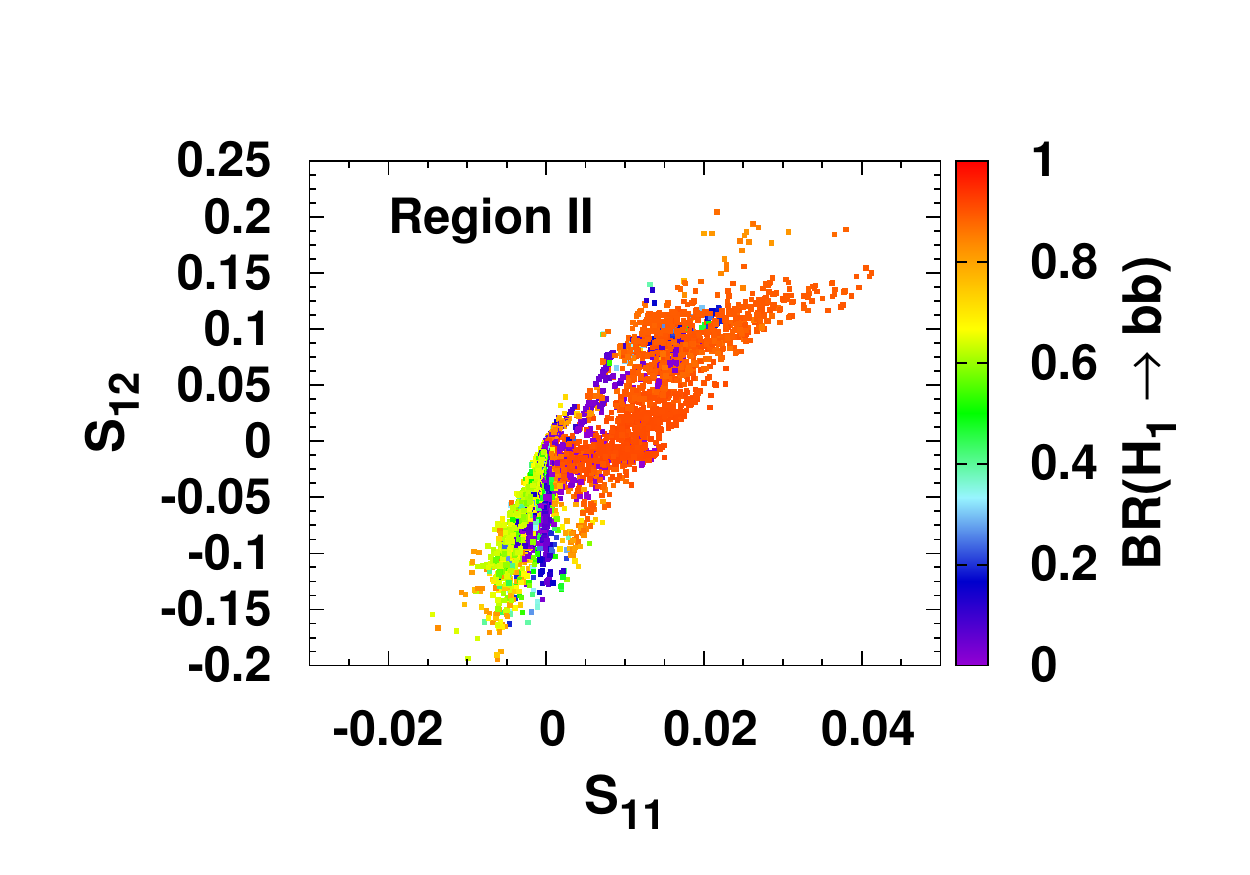}

\caption[]{ 
Left: The separation of Region I/II (dark/light grey (blue) dots) can be easily observed in the Higgs mixing matrix element plane of $S_{11}$ and $S_{12}$. In Region I the value of $S_{11}$ is positive, while in Region II $S_{11}$ can have negative and positive values. This change of sign leads to small values of $S_{11}$. 
Right: 
This is a cutout of the left panel for $S_{11}$ close to zero, showing as shaded (color) coding the strong variation of the light Higgs branching
ratio into b-quarks in the region where $S_{11}$ changes sign. For this small region the branching ratio in c-quarks becomes dominant.
}
\label{f1}
\end{center}
\end{figure}

\begin{figure}
\footnotesize{\bf{Region I} (large $\lambda,\kappa$, small $\tan\beta$) \hspace{1.5cm} \bf{Region II} (small $\lambda,\kappa$, large $\tan\beta$)} 
\begin{center}
\includegraphics[width=0.49\textwidth]{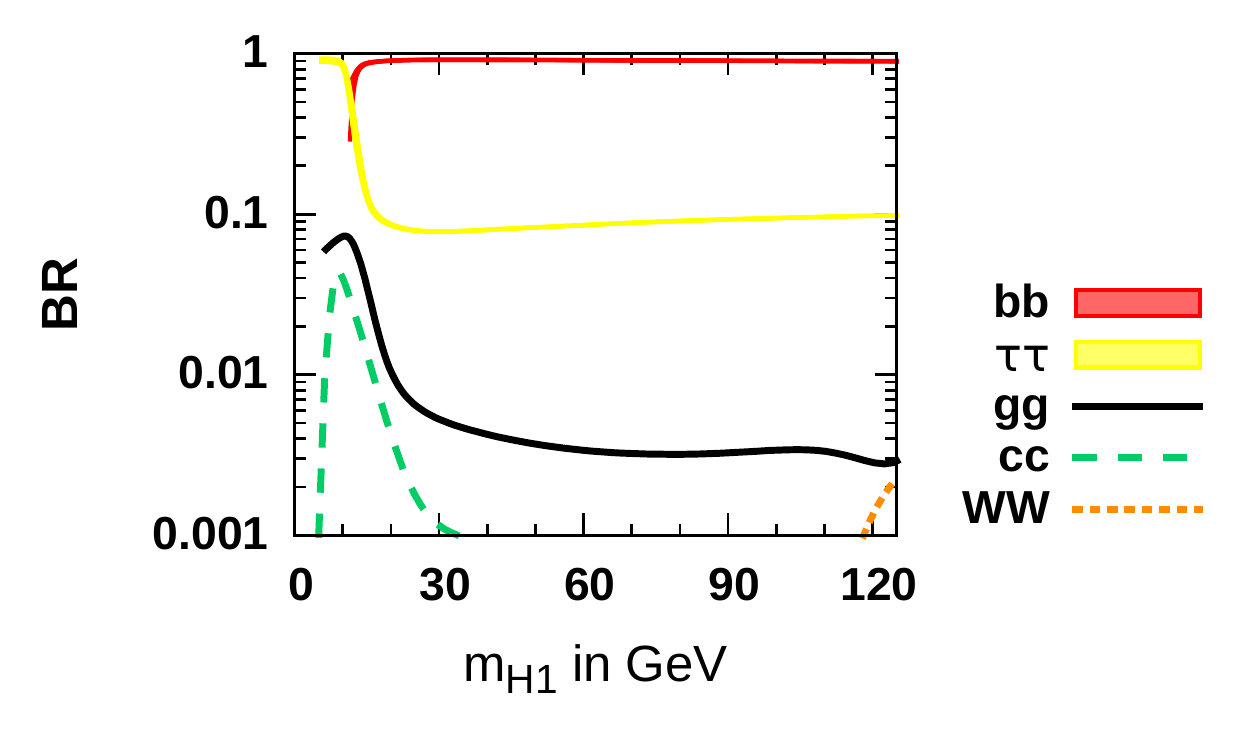}
\includegraphics[width=0.49\textwidth]{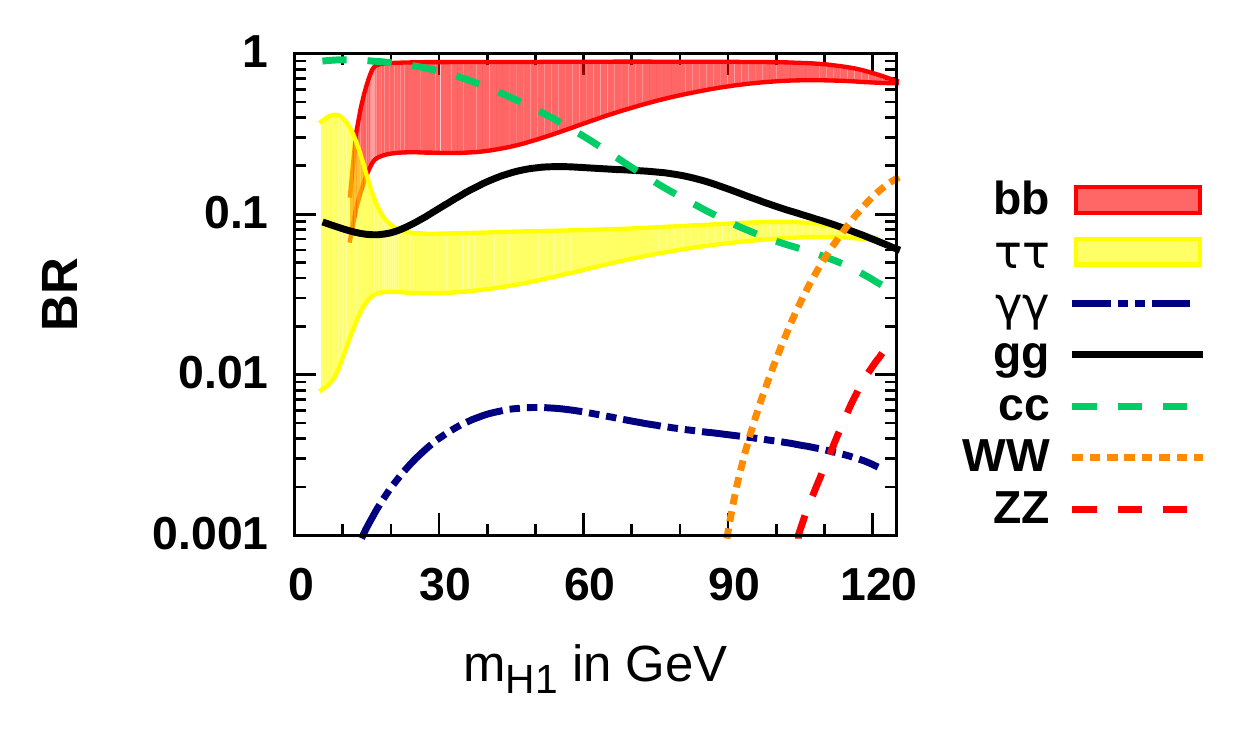}
\caption[]{
Branching ratios of the light Higgs boson $H_1$ for Region I and II (left and right), respectively. The allowed range of the branching ratios for $bb$ and $\tau\tau$ is shown by the shaded (colored) bands, which includes 68\% of the sampled points around the most probable branching ratio with 34\% of the sampled points on each side. In the few cases, where less than 34\% of the points are on one side, the other side of the interval is chosen larger such that the whole area has a 68\% probability.   
The other branching ratios are represented by lines in order not to clutter the figure. The lines represent the most probable branching ratio. 
In Region I the bands for the branching ratios into $bb$ and $\tau\tau$ are very narrow. In Region II, the decay into b-quarks can become small, which leads to a broad allowed band for the corresponding branching ratio (see text). All other branching ratios represented by a line in Region II are surrounded by a broad band as well. The sum of the branching ratios adds up to 1 and can include decays into light pseudo-scalar Higgs bosons $A_1$ and neutralinos, if kinematically allowed. Such cases are exemplified in the benchmark points as given in the Appendix.
}
\label{f3}
\end{center}
\end{figure}

\section{A light Higgs boson below 125 GeV in the NMSSM}
\label{br-higgs}

As already discussed in Sect. \ref{higgs}, the mixing matrix elements of the lightest Higgs boson $S_{11}$ and $S_{12}$ determine the couplings to the b- and t-quarks (see Eq. \ref{coupling}). 
The values of $S_{11}$ and $S_{12}$ are determined by the fitted 7 parameters from the upper right box in Fig. \ref{method} and are shown on the left-hand side of Fig. \ref{f1} for Regions I (dark grey (dark blue) dots) and II (light grey (light blue) dots). The range of $S_{12}$ is similar for both regions, but the range of $S_{11}$ differs. One observes that $S_{11}$ is always small with positive and negative values around $S_{11}=0$ for Region II (light grey (light blue) points).
This means that the coupling to b-quarks, and hence the branching ratio, goes through zero as indicated by the shaded (color) coding on the right-hand side of Fig. \ref{f1}. In this case the branching ratios to other channels like $gg$,$c\bar{c},...$, increase correspondingly, as shown in Fig. \ref{f3} for Region I and II (left and right), respectively. 
The region with $S_{11} \sim 0$ corresponds only to a small part of the parameter space as demonstrated on the right-hand side of Fig. \ref{f1} by the dark blue points. 
For Region I, $S_{11}$ is always positive, so no regions with branching ratios to b-quarks close to zero occur and the branching ratio into b-quarks always dominates, if kinematically allowed.
For Higgs boson masses below the b-quark threshold the decay into tau leptons increases, as shown in Fig. \ref{f3}. In both regions the branching ratio into b-quarks can be reduced if the decay into neutralinos or light pseudo-scalar Higgs bosons is kinematically allowed, which will be discussed in more detail in the following section.

\begin{figure}
\begin{center}
\hspace{-2.5cm}

\begin{minipage}{\textwidth}
\begin{tabular}{ccc}
&  \multirow{8}{*}{\includegraphics[width=0.44\textwidth]{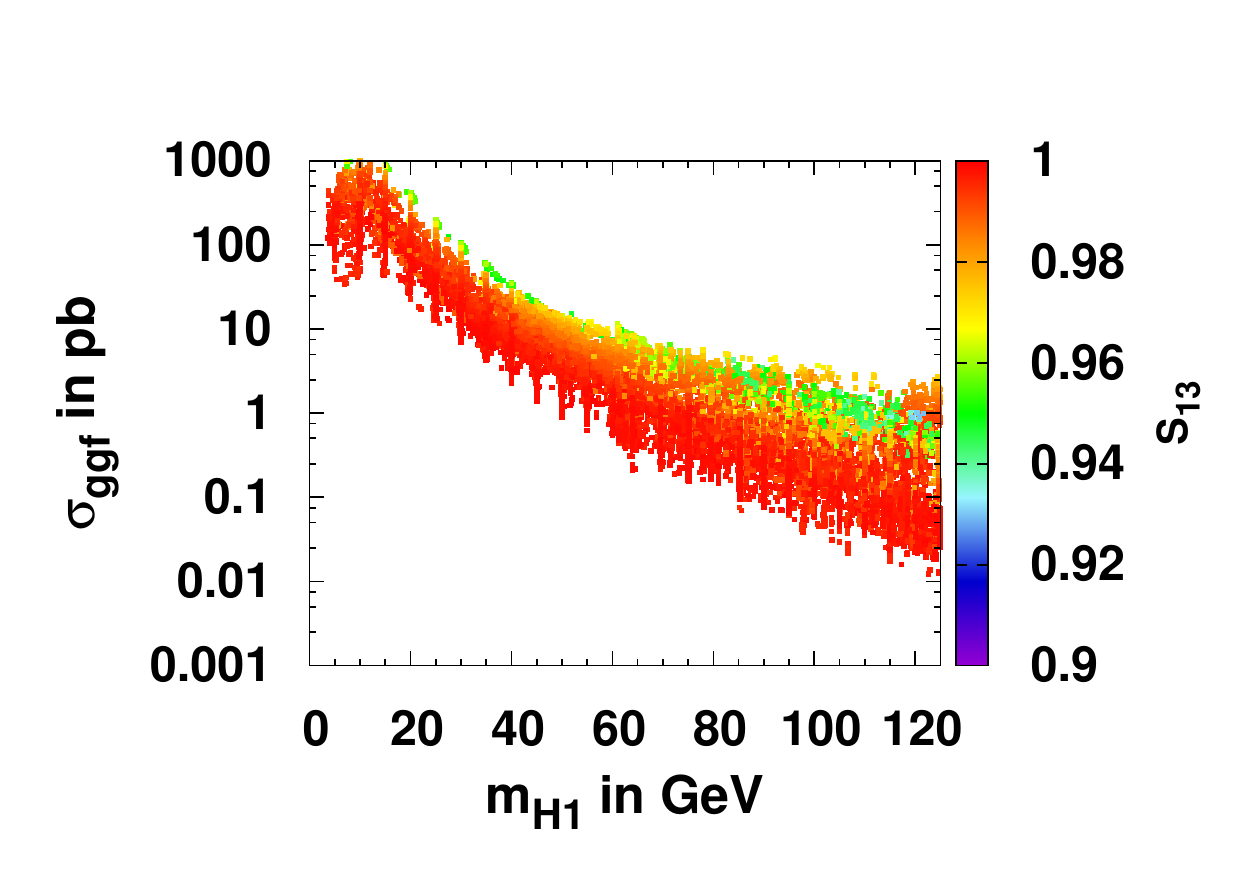}} & \multirow{9}{*}{\includegraphics[width=0.44\textwidth]{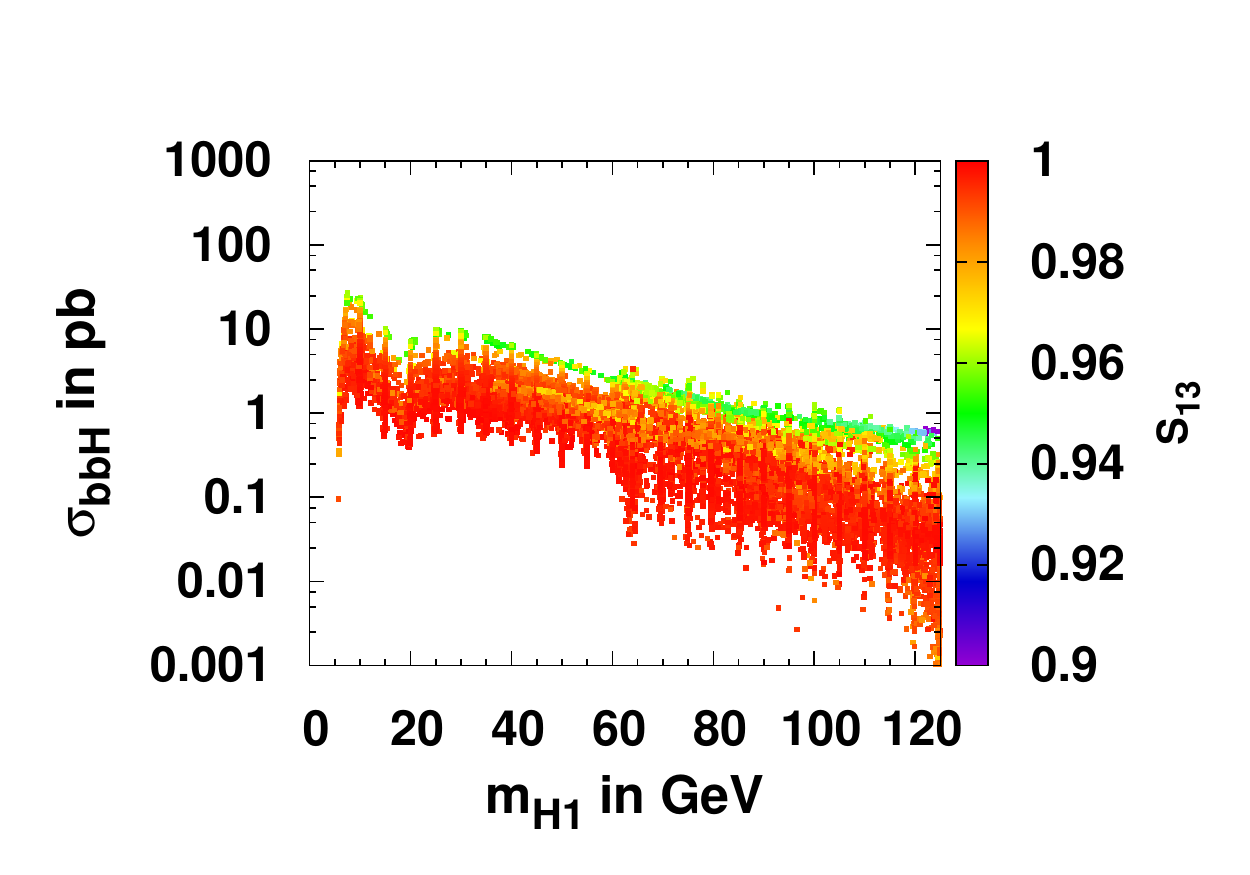}} \\
 &  & \\
 &  & \\
\footnotesize{\bf{Region I}} & & \\
\scriptsize{(large $\lambda,\kappa$, } &  & \\
\scriptsize{small $\tan\beta$)} &  & \\
 &  & \\
 &  & \\
&  \multirow{8}{*}{\includegraphics[width=0.44\textwidth]{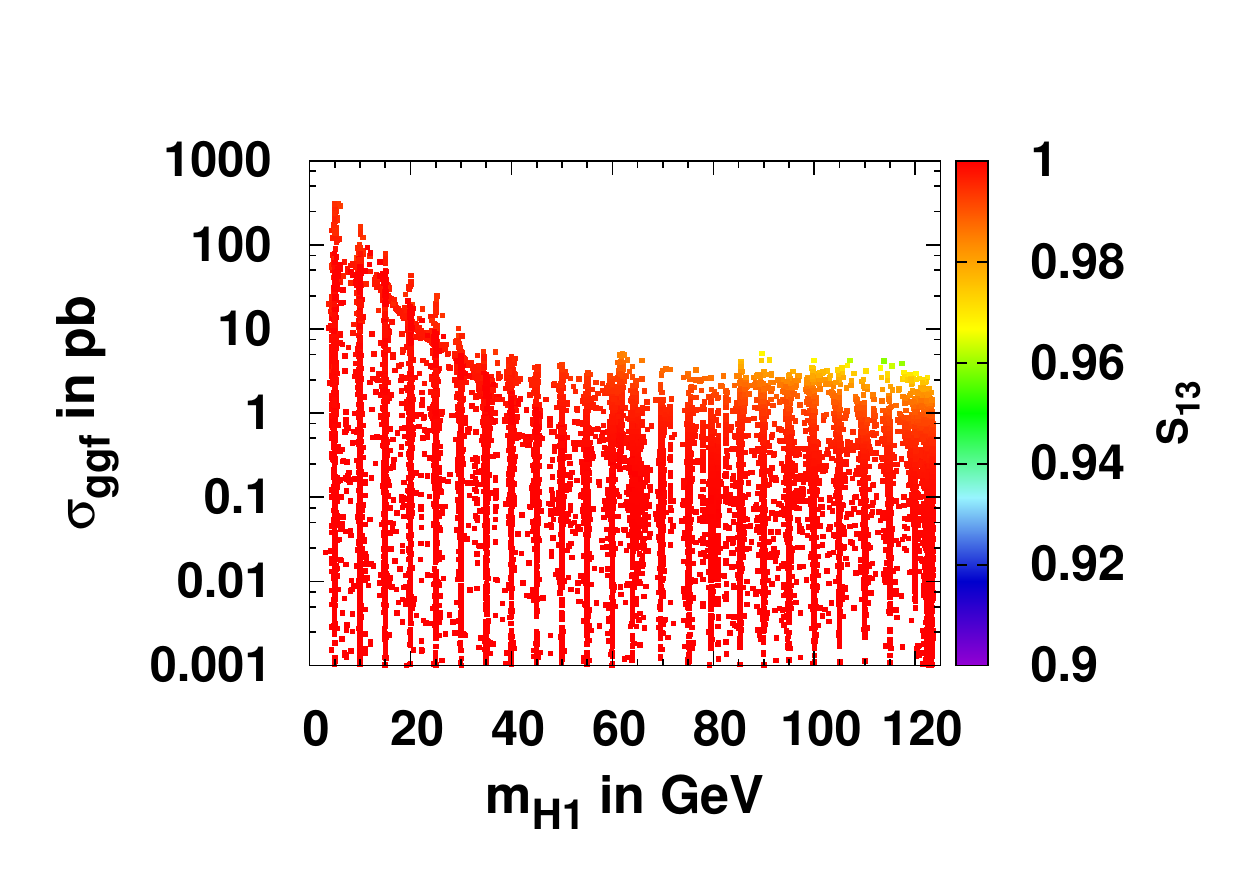}} & \multirow{9}{*}{\includegraphics[width=0.44\textwidth]{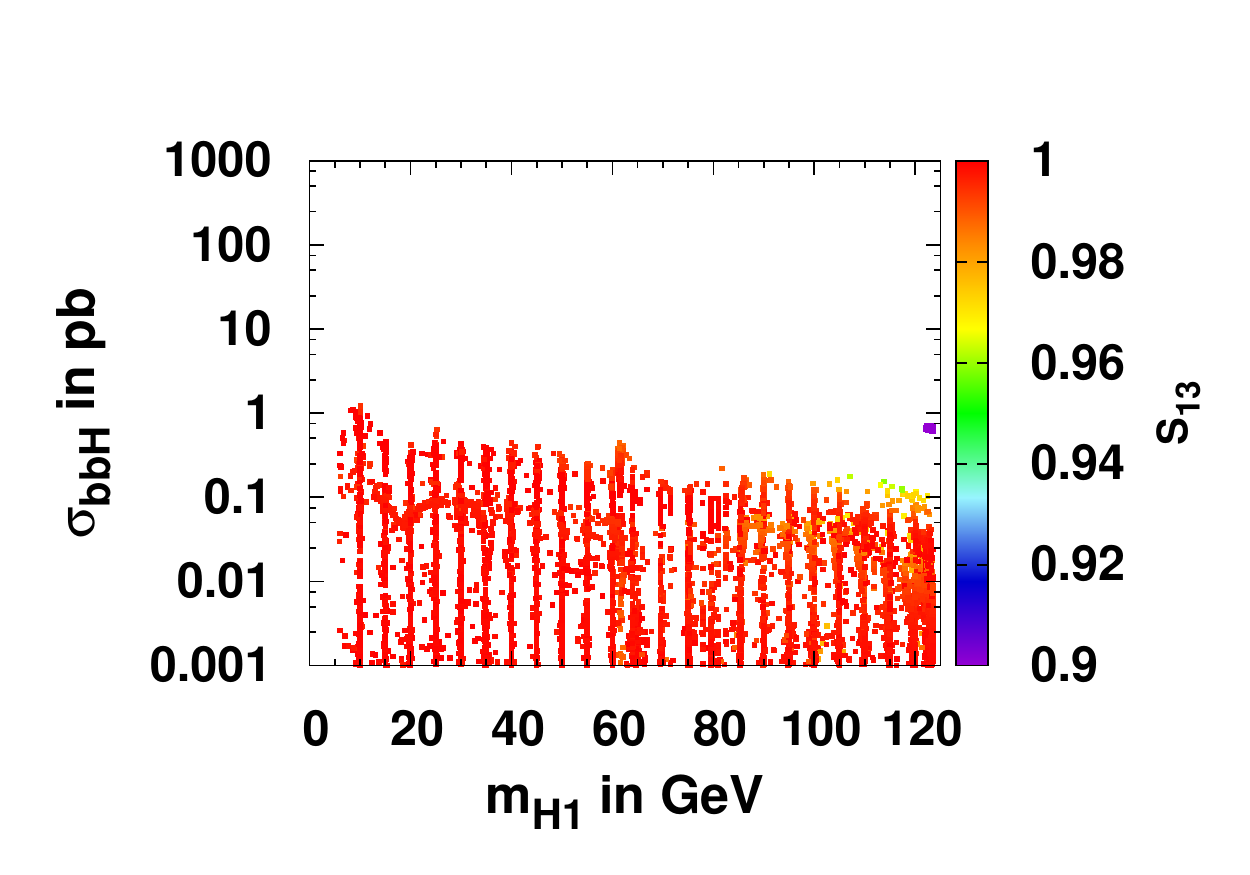}} \\
 &  & \\
 &  & \\
\footnotesize{\bf{Region II}} & & \\
\scriptsize{(small $\lambda,\kappa$, } &  & \\
\scriptsize{large $\tan\beta$)}  &  & \\
 &  & \\
 &  & \\
\end{tabular}
\end{minipage}
\caption[]{Production cross section in pb versus light Higgs boson mass $m_{H1}$ via gluon fusion (left) and in association with b-quarks (right) for Region I/II (top/bottom). The shaded (color) coding corresponds to the singlet content $S_{13}$ of the lightest Higgs boson.
}
\label{f0}
\end{center}
\end{figure}

If the coupling to b-quarks can become zero, the associated production cross section can become zero as well, which leads to a large variation in the cross section as shown in the lower panel of Fig. \ref{f0} for Region II. Here the two main production processes for Higgs production at the LHC,  via gluon fusion (ggf) or via the production in association with b-quarks (bbH), are shown. 
The cross sections are small, since $H_1$ is largely a singlet, as represented by the shaded (color) coding in Fig. \ref{f0} indicating that $S_{13}$ (see Eq. \ref{eq1}) is usually above 0.95. One observes that in both regions the gluon fusion production is dominant in spite of the fact that the associated production has the "$\tan\beta$-enhancement", i.e. $\propto \tan^2\beta$, which is important in Region II. But the associated production is also proportional to the Higgs coupling to b-quarks, i.e. $\propto S_{11}^2$, which can be small in Region II (see Fig. \ref{f1} right-hand side), thus leading to a small cross section even at large values of $\tan\beta$.

\begin{figure}
\begin{center}
\hspace{-2.5cm}
\begin{minipage}{\textwidth}
\begin{tabular}{ccc}
&  \multirow{8}{*}{\includegraphics[width=0.44\textwidth]{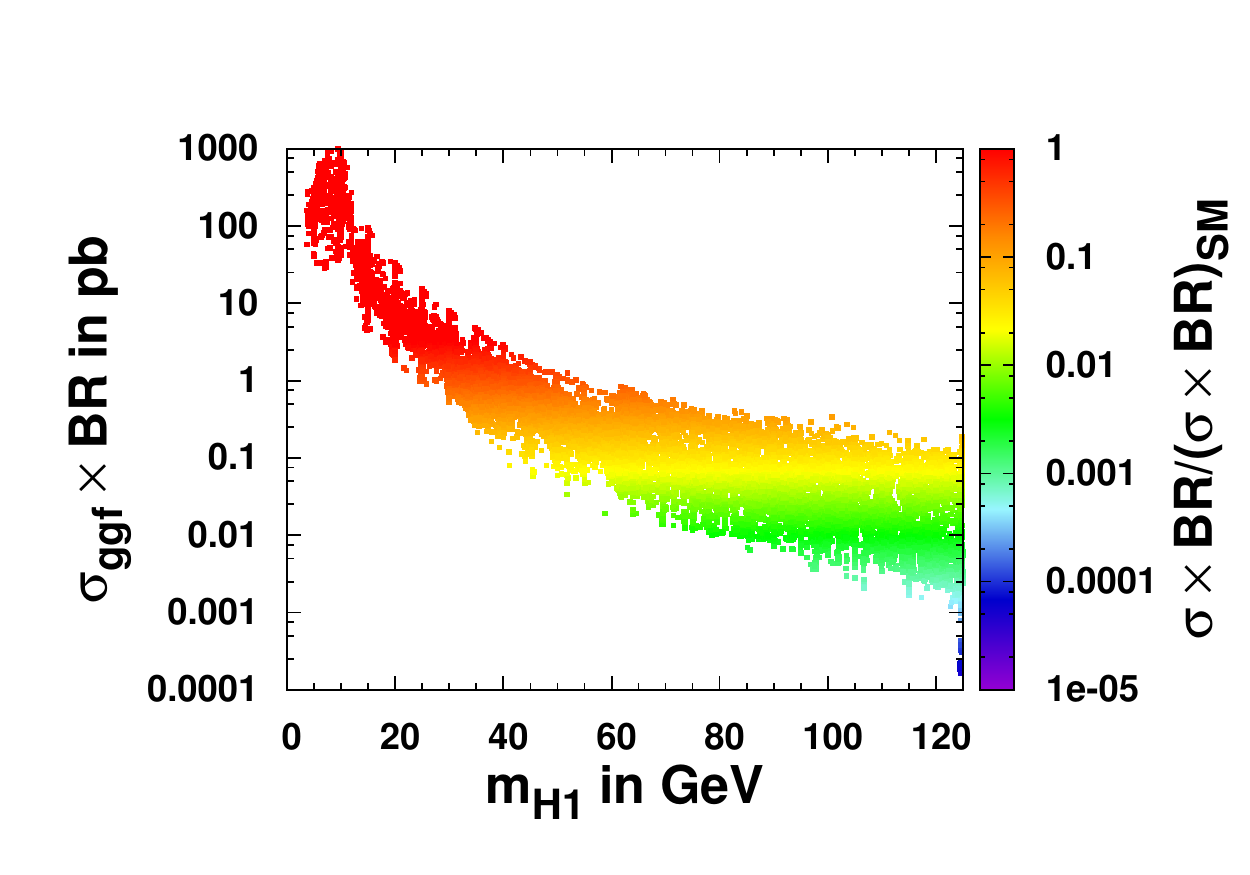}} & \multirow{9}{*}{\includegraphics[width=0.44\textwidth]{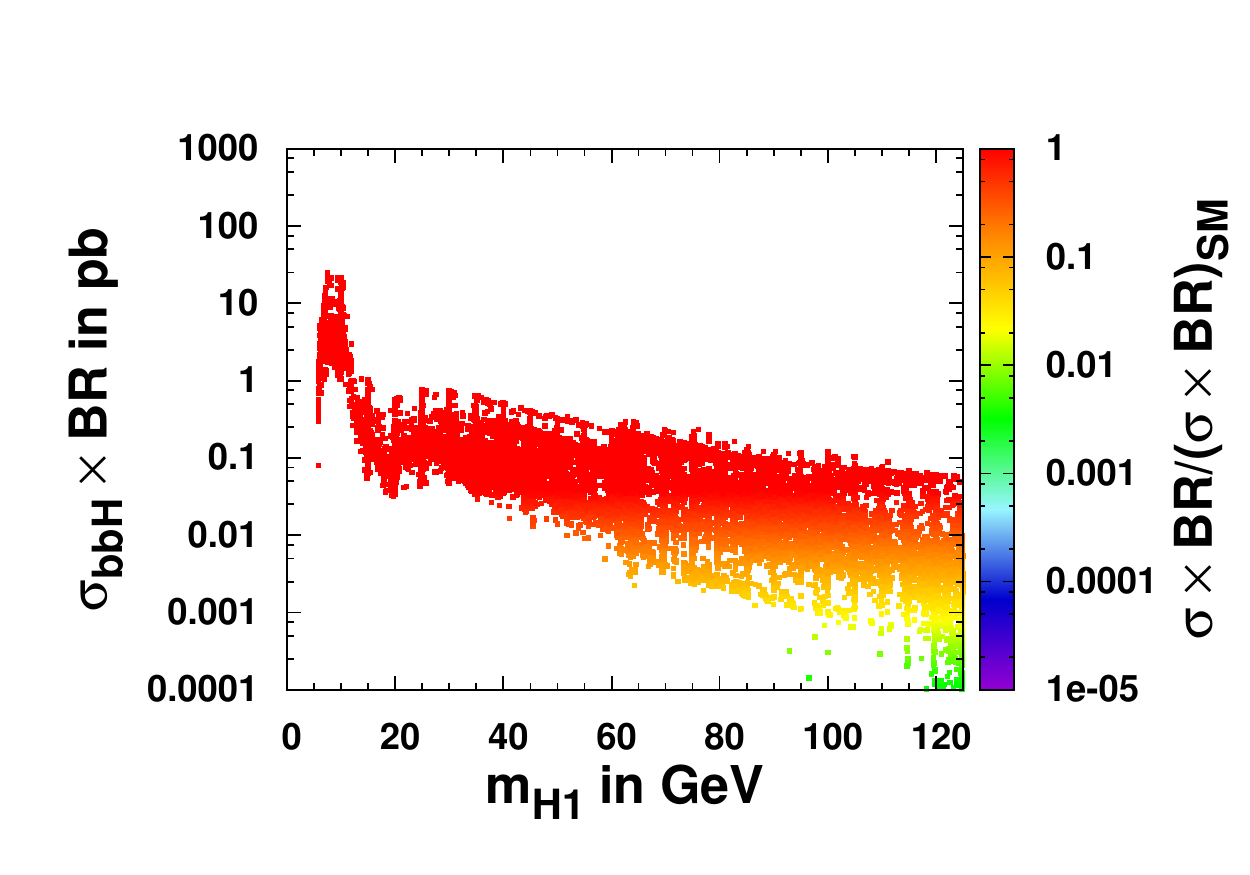}} \\
 &  & \\
 &  & \\
\footnotesize{\bf{Region I}} & & \\
\scriptsize{(large $\lambda,\kappa$, } &  & \\
\scriptsize{small $\tan\beta$)} &  & \\
 &  & \\
 &  & \\
&  \multirow{8}{*}{\includegraphics[width=0.44\textwidth]{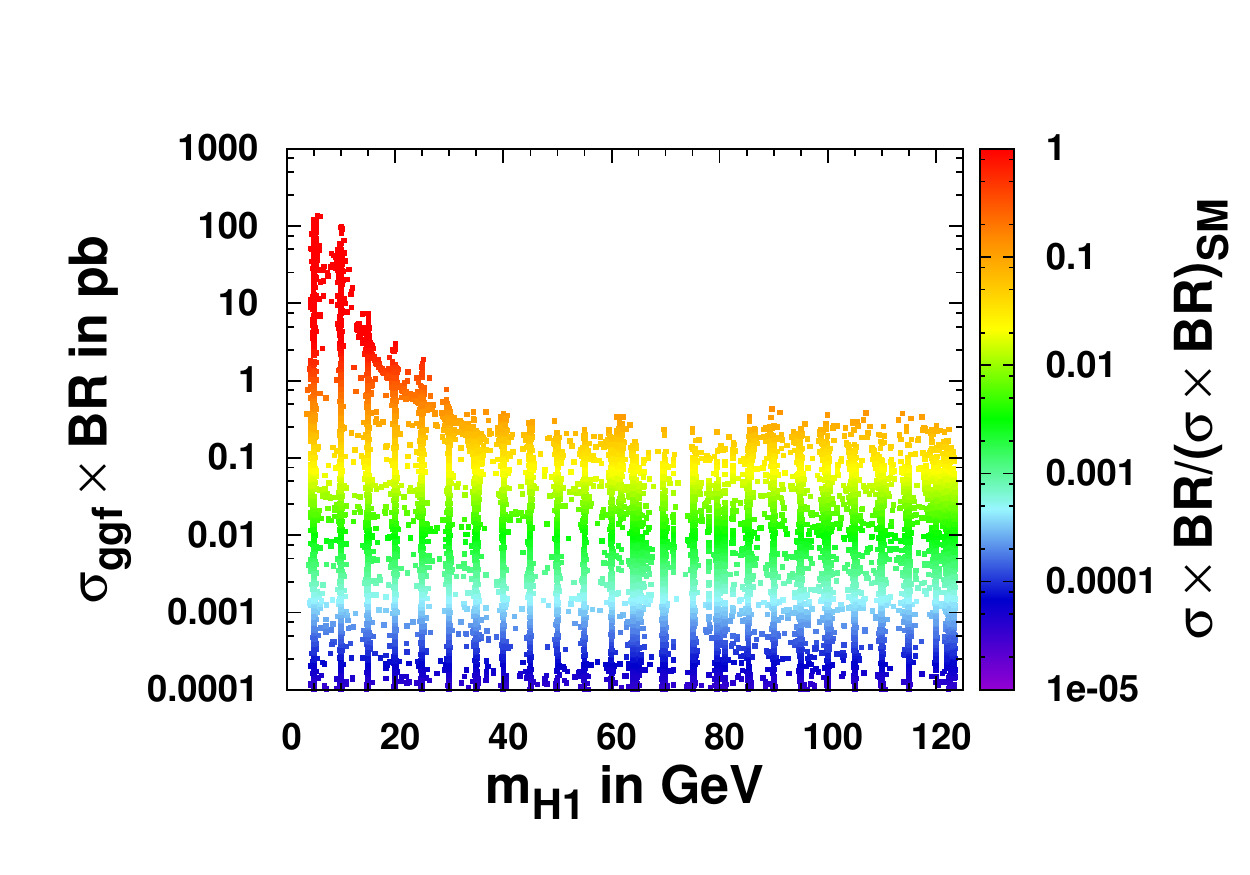}} & \multirow{9}{*}{\includegraphics[width=0.44\textwidth]{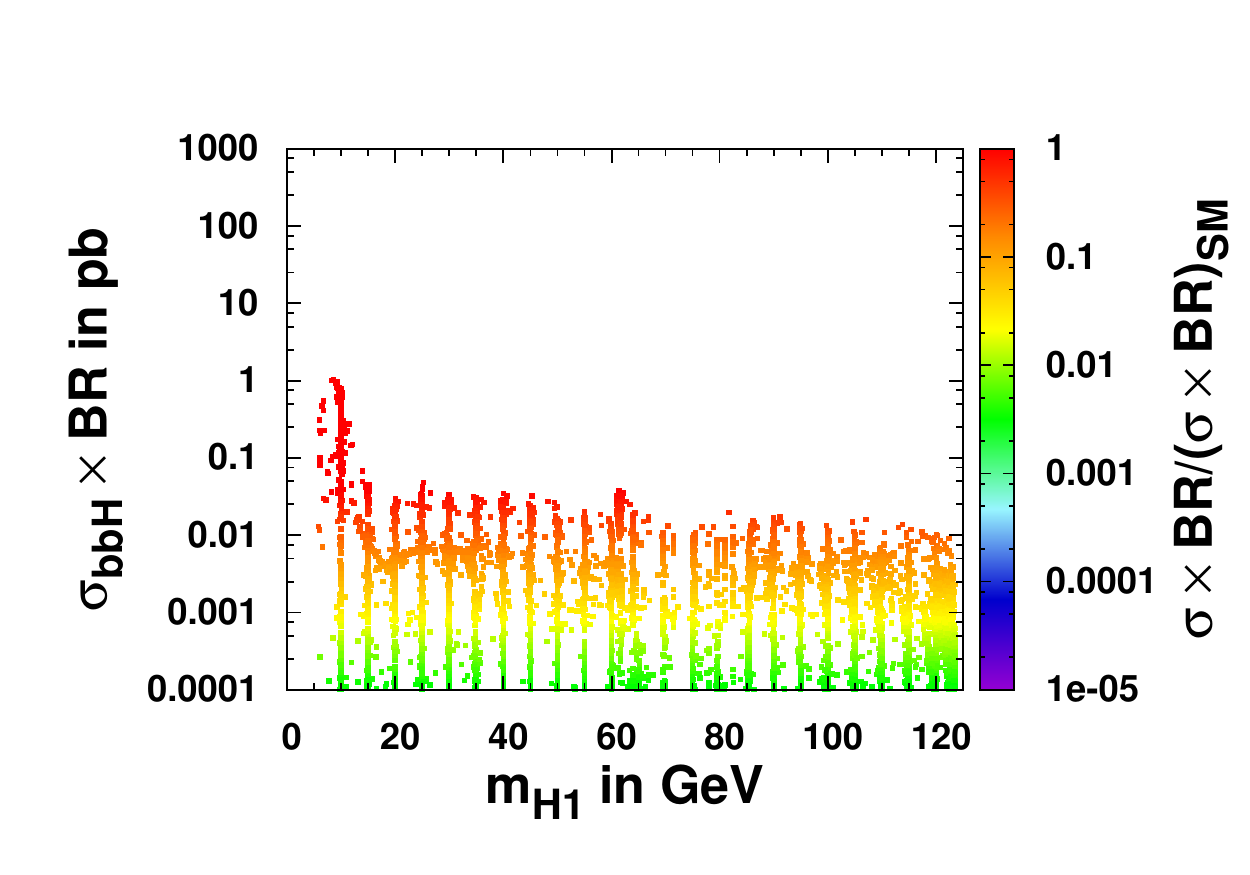}} \\
 &  & \\
 &  & \\
\footnotesize{\bf{Region II}} & & \\
\scriptsize{(small $\lambda,\kappa$, } &  & \\
\scriptsize{large $\tan\beta$)}  &  & \\
 &  & \\
 &  & \\
\end{tabular}
\end{minipage}
\caption[]{Cross section times branching ratio via gluon fusion (left) and in association with b-quarks (right) for Region I/II (top/bottom) into tau final states. The ratio of $\sigma \times BR / (\sigma \times BR)_{SM}$ is indicated by the shaded (color) coding.  
}
\label{f3-0}
\end{center}
\end{figure}

\begin{table}
\tiny
\centering
\caption{Summary of the lightest Higgs boson $H_1$ branching ratios (in \%) and cross section times branching ratio in fb for 14 TeV for different $m_{H1}$ mass ranges in GeV in Region I. The interval includes 68\% of the sampled points around the most probable value of the corresponding branching ratio. Values below 0.01 are set to zero. For comparison of the discovery potential the cross section times branching ratio for the 125 GeV Higgs boson are for the ggf channel $\gamma\gamma: 112$ fb,$~ZZ: 1321$ fb,$~Z\gamma: 76$ fb,$~\tau\tau: 3090$ fb and for the bbH channel $\gamma\gamma: 1.3$ fb,$~ZZ: 15.4$ fb,$~Z\gamma: 0.8$ fb,$~\tau\tau: 36$ fb. 
\label{t5}}
\begin{tabular}{l|c|c|c||c|c|c}
	\hline\noalign{\smallskip}
	& \multicolumn{3}{c||}{0-20 GeV} & \multicolumn{3}{c}{20-40 GeV} \\  
	\noalign{\smallskip}\hline\noalign{\smallskip}
	Name & BR in \% & $\sigma_{ggf} \times BR $ in fb & $\sigma_{bbH} \times BR $ in fb &  BR in \% & $\sigma_{ggf} \times BR $ in fb & $\sigma_{bbH} \times BR $ in fb\\
	\noalign{\smallskip}\hline\noalign{\smallskip}
	$\tau \tau$ & 10.1 - 92.8    &  12960.8 - 374228.8  & 34.6 - 2283.0 & 7.7 - 8.0   &  585.9 - 3782.3  &   61.4 -  202.2      \\
	$\gamma \gamma$ &  $<0.01$ & 0.62 - 13.7 & 0.0 - 0.07 &  $<0.01$ & 0.02 - 0.22 & 0.0 - 0.01  \\ 
	$Z \gamma $ & - & - & -&     - & - & -\\ 
	$Z Z$ & - & -  & -    & -& -&-\\
	$W W$ & - & -  & -    & -& -&-\\
	$A_1 A_1$ & - & -     & - & 20.3 -67.0 & 336.6 - 16157.7 & 189.4 - 995.7  \\ 
	$\tilde{\chi}_0^1  \tilde{\chi}_0^1$&  -   & -   & - & 0.0 - 35.3 & 0.0 - 5329.2 & 231.1 - 392.4  \\ 
	$b b$ & 1.8 - 89.4 & 0.0 - 819769.8 &  833.9 - 2219.1 & 91.5 - 91.7  & 6715.2 - 43022.2 &  717.4 - 2387.2\\
	$c c$ & 0.08 - 0.93 &  22.7 - 3053.7 &  0.35 - 36.2  & 0.07 - 0.22 & 9.1 - 80.1 & 0.94 - 4.3 \\
	$g g$ & 0.98 - 8.1 &  1267.1 - 31858.6 &  12.5 - 435.8 & 0.43 - 0.64  & 33.7 - 296.3 & 3.7 -  12.7\\
	\noalign{\smallskip}\hline\noalign{\smallskip}
	\noalign{\smallskip}\hline\noalign{\smallskip}
	& \multicolumn{3}{c||}{40-90 GeV} & \multicolumn{3}{c}{90-120 GeV} \\ 
	\noalign{\smallskip}\hline\noalign{\smallskip}
	Name & BR in \% & $\sigma_{ggf} \times BR $& $\sigma_{bbH} \times BR $ &  BR in \% & $\sigma_{ggf} \times BR $& $\sigma_{bbh} \times BR $\\
	\noalign{\smallskip}\hline\noalign{\smallskip}
	$\tau \tau$ &  8.2 - 9.1  &  47.3 - 520.2 &  18.4 - 163.0  & 9.5 - 9.9 &  6.2 - 73.4 & 3.6 - 54.7   \\
	$\gamma \gamma$ &  $<0.01$ & 0.01 - 0.13 & 0.0 - 0.05 & 0.0 - 0.02 & 0.0 - 0.04 & 0.0 - 0.03 \\ 
	$Z \gamma $ & - &     - & - & $<0.01$ & $<0.01$  & $<0.01$ \\ 
	$Z Z$ & - & -  & -    &  0.0 - 0.01 & $<0.01$ & $<0.01$ \\
	$W W$ & 0.0 - 0.01 &  $<0.01$  &  $<0.01$    & 0.0 - 0.09  & 0.0 - 0.05 & 0.0 - 0.04 \\
	$A_1 A_1$ & 65.5 - 95.4      & 256.5 - 12568.1 &  294.5 - 745.0 & -&- &-\\ 
	$\tilde{\chi}_0^1  \tilde{\chi}_0^1$& 0.0 - 64.8  & 0.0 - 1176.9  &   67.6 - 210.5 & -  &-  &-\\ 
	$b b$ & 90.7 - 91.5  & 473.4 - 5788.6 &  190.7 - 1721.4 & 89.7 - 90.1 & 58.0 - 706.6 & 33.1 - 519.2 \\
	$c c$ & 0.04 - 0.11  & 0.03 - 2.8 &  0.15 - 1.9 & 0.03 - 0.11 & 0.0 - 0.19  &  0.0 - 0.26 \\
	$g g$ & 0.26 - 0.36 & 1.6 - 21.7  & 0.69 - 6.7  & 0.21 - 0.40  & 0.19 - 2.3 & 0.12 - 1.6 \\
	\noalign{\smallskip}\hline
\end{tabular}
\end{table}

\begin{table}
\tiny
\centering
\caption{Same table as Table \ref{t5} but for Region II. \label{t6}}
\begin{tabular}{l|c|c|c||c|c|c}
	\hline\noalign{\smallskip}
	& \multicolumn{3}{c||}{0-20 GeV} & \multicolumn{3}{c}{20-40 GeV} \\  
	\noalign{\smallskip}\hline\noalign{\smallskip}
	Name & BR in \% & $\sigma_{ggf} \times BR $ in fb & $\sigma_{bbH} \times BR $ in fb &  BR in \% & $\sigma_{ggf} \times BR $ in fb& $\sigma_{bbH} \times BR $ in fb\\
	\noalign{\smallskip}\hline\noalign{\smallskip}
	$\tau \tau$ & 1.2 - 51.0 & 0.0 - 2700.8 & 0.0 - 20.8  &  2.1 - 7.9 & 0.0 - 60.8 &  0.0 - 0.68 \\
	$\gamma \gamma$ & 0.0 - 0.08 & 0.0 - 0.65 & $<0.01$ & 0.0 - 0.42  & 0.0 - 0.50 &  $<0.01$ \\ 
	$Z \gamma $ & -  & -     &-  & -& -&-\\ 
	$Z Z$ & - & -  & -    & -& -&-\\
	$W W$ & - & -  & -    & -& -&-\\
	$A_1 A_1$ &  46.7 - 96.7 & 0.0 - 18904.0  & 0.0 - 117.2 & 79.7 - 97.6  & 0.0 - 2867.8 & 0.0 - 184.0 \\ 
	$\tilde{\chi}_0^1  \tilde{\chi}_0^1$& -    &  - &  -& -& -&-\\ 
	$b b$ & 0.0 - 89.4  &  0.0 - 48602.7 &  0.0 - 374.8  & 24.5 - 90.1      &  0.0 - 693.6  &  0.0 - 8.1 \\
	$c c$ & 33.7 - 93.9  & 0.0 - 2975.7 & 0.0 - 5.0 & 3.1 - 77.2  & 0.0 - 109.5 & 0.0 - 0.21 \\
	$g g$ & 2.6 - 10.2 & 0.0 - 694.3  & 0.0 - 2.3  & 1.2 - 15.2      & 0.0 - 36.3 & 0.0 - 0.20 \\
	\noalign{\smallskip}\hline\noalign{\smallskip}
	\noalign{\smallskip}\hline\noalign{\smallskip}
	& \multicolumn{3}{c||}{40-90 GeV} & \multicolumn{3}{c}{90-120 GeV} \\ 
	\noalign{\smallskip}\hline\noalign{\smallskip}
	Name & BR in \% & $\sigma_{ggf} \times BR $& $\sigma_{bbH} \times BR $ &  BR in \% & $\sigma_{ggf} \times BR $& $\sigma_{bbH} \times BR $\\
	\noalign{\smallskip}\hline\noalign{\smallskip}
	$\tau \tau$ &   4.2 - 9.0 & 0.0 - 7.1  &  0.0 - 0.65  & 6.9 - 9.2 & 0.0 - 5.5 &  0.0 - 0.29 \\
	$\gamma \gamma$ & 0.02 - 0.93  & 0.0 - 0.95  & $<0.01$ & 0.16 - 2.1  & 0.0 - 0.92  & $<0.01$  \\ 
	$Z \gamma $ & - & -     &-  & 0.0 - 0.06 &  0.0 - 0.04 & $<0.01$\\ 
	$Z Z$ &  -& -  & -    & 0.0 -  0.28  & 0.0 - 0.26 &  $<0.01$\\
	$W W$ & 0.0 - 0.17 &  0.0 - 0.53 &  0.0 - 0.01 & 0.40 - 11.1   &  0.0 - 3.8 & 0.0 - 0.03 \\
	$A_1 A_1$ &  95.9 - 98.8 &  0.0 - 2462.9 & 0.0 - 217.9 &  71.5 - 98.1  & 0.0 - 426.5 & 0.0 - 43.5 \\ 
	$\tilde{\chi}_0^1  \tilde{\chi}_0^1$& -&- & -   &-   &  -&-\\ 
	$b b$ &  46.2 - 90.2 & 0.0 - 80.8  &  0.0 - 7.3 & 68.4 - 89.6  & 0.0 - 52.5 & 0.0 - 2.8 \\
	$c c$ & 0.86 - 49.6  & 0.0 - 28.8  &  0.0 - 0.06  & 0.74 - 9.2 & 0.0 - 13.4 & 0.0 - 0.04 \\
	$g g$ & 0.83 - 28.7  & 0.0 - 26.6 & 0.0 - 0.05 & 4.1 - 51.1 & 0.0 - 21.2 & 0.0 - 0.05 \\ 
	\noalign{\smallskip}\hline
\end{tabular}
\end{table}

\section{Discovery potential for selected final states}

From Fig. \ref{f3} it is clear that many signatures are possible: $\gamma\gamma,\tau\tau,WW,$ $ZZ,bb,cc,gg,...$ over a large range of the $H_1$ mass. It is beyond the scope of this letter to discuss quantitatively the discovery potential of each of these channels, since for low masses the background rapidly increases and efficiencies decrease, so the discovery potential for each channel can only be obtained from a detailed simulation. Benchmark points for such detailed simulations can be found in the Appendix in Tables \ref{t41} and \ref{t42} for each discovery channel. 
Instead, we compare the cross sections times branching ratios with the corresponding value for the observed Higgs boson in order to get a feeling for the observability. 

Most of the dominant final states shown in Fig. \ref{f3} include quarks and gluons, which lead to multiple jets in the detector. Those final states are challenging because of the large hadronic background at a hadron collider. 
In both regions the most promising discovery channel is the decay into tau leptons. The corresponding cross sections times branching ratios for 14 TeV are shown in Fig. \ref{f3-0} for Region I (top) and II (bottom) for both production modes. The shaded (color) coding represents the value of the cross section times branching ratio normalized to the SM value $\sigma \times BR/(\sigma \times BR)_{SM}$, where the denominator refers to the SM values for the 125 GeV SM Higgs boson decaying into tau leptons. 
The future integrated luminosity is assumed to reach 300 (3000) fb$^{-1}$, which is one (two) orders of magnitude higher than the luminosity from the observation of the SM-like Higgs boson into tau leptons, see Ref. \cite{Sirunyan:2017khh}. Assuming that the discovery potential scales with the luminosity L as $\sqrt{L}$ and the efficiency stays constant implies that the red (orange) areas in Fig. \ref{f3-0} are of interest to look at for L=300 (3000) fb$^{-1}$, if similar efficiencies and backgrounds are assumed. 
The results of Fig. \ref{f3-0}, together with other decay channels, have been summarized in Tables \ref{t5} and \ref{t6} for Region I and II and various $H_1$ mass ranges for 68\% of the sampled points around the most probable value, following the recipe as detailed in the caption of Fig. \ref{f3}. Except for the $H_1$ mass region below 20 GeV the cross section times branching ratio into tau leptons is of the order of a few hundreds of fb for both regions. 
We select benchmark points with the maximal cross section times branching ratio from the range of branching ratios in Tables \ref{t5} and \ref{t6}. The benchmark points have been defined in Tables \ref{t41} and \ref{t42} of the Appendix, including mass information and NMSSMTools parameters in Table \ref{t43}. 
  
Decays including $Z(W)$ bosons in the final state can only be used to access a mass range of $H_1$ above 90(80) GeV, respectively, as can be seen from Tables \ref{t5} and \ref{t6}. 
The absolute values of the cross sections times branching ratios into Z-bosons are small, but the corresponding values for the W-boson in Region II can be larger, if kinematically allowed. 
However, the neutrinos in the decay of the W-boson broaden the mass peaks, thus reducing the sensitivity in comparison with the $ZZ$ and $\gamma\gamma$ final states.

The $\gamma\gamma$ final states do not suffer from kinematic limits like the $Z$ and $W$ final states, so this final state can be used to access the whole $H_1$ mass range, as can be seen from Tables \ref{t5} and \ref{t6}. The decay via a top loop increases for large couplings to up-type fermions, which is the case if $S_{11} \sim 0$ in Region II. This leads to large cross sections times branching ratios for $\gamma\gamma$ final states up to 10 fb for the $H_1$ mass region above 40 GeV in Region II. The cross section times branching ratio of the order of 1 fb shown in Table \ref{t6} corresponds to the interval including 68\% of the sampled points. Furthermore, the normalized cross section times branching ratio $\sigma \times BR/(\sigma \times BR)_{SM}$ is above $10\%$ for the whole mass range. 

Besides final states including SM particles new decays into pairs of the lightest pseudo-scalar Higgs boson $A_1$ or the lightest neutralino $\tilde{\chi}_1^0$ are possible in the NMSSM. 
However, this decay is possible in the small region of the parameter space with $m_{A1}/m_{\tilde{\chi}_1^0}< 0.5 m_{H1}$. The mass of $A_1$ and $\tilde{\chi}_1^0$ are correlated, so both signatures happen in the same region of parameter space. 
The decay into the light pseudo-scalar Higgs boson $A_1$ is possible in both regions, but the decay into neutralinos is only possible in Region I. The reason is simply the small values of $\lambda$ and $\kappa$ in Region II, which lead to small mixing in the Higgs and neutralino sectors. In this case the $H_1$ mass and neutralino mass can be approximated by   $m_{H1} < (2\kappa/\lambda) \cdot \mu_{eff}$ and $ m_{\tilde{\chi}_0^1} \sim (2\kappa/\lambda) \cdot \mu_{eff}$, so the decay $H_1 \rightarrow \tilde{\chi}_1^0 \tilde{\chi}_1^0$ is kinematically suppressed.   
If kinematically allowed, those decays can dominate and reach branching ratios up to 90-100\%, as shown Tables \ref{t5} and \ref{t6}.
Neutralino final states will give events with large missing transverse energy (MET), while the light pseudo-scalar Higgs boson will further decay predominantly into b-quarks and tau leptons, leading to $bb\tau\tau$ or $4\tau$ final states.

\section{Conclusion}
We surveyed the branching ratios of the singlet-like Higgs boson below 125 GeV in the NMSSM in two regions, one for low and one for large va\-lues of the couplings $\lambda,\kappa$. 
From the branching ratios we consider the following channels to be the most promising for future searches for a singlet-like NMSSM Higgs boson with a mass below 125 GeV at the LHC: $\tau\tau, \gamma\gamma, Z\gamma, ZZ, WW, A_1 A_1$ and $\tilde{\chi}_1^0 \tilde{\chi}_1^0$. 
We compare the cross sections times branching ratios with the corresponding value for the observed 125 GeV Higgs boson in order to get a feeling for the observability for the two dominant Higgs production modes (gluon fusion and associated production with b-quarks) in Tables \ref{t5} and \ref{t6} for both regions. Selected benchmark points for all discovery channels have been given in Tables \ref{t41} and \ref{t42} in the Appendix for a quantitative determination of the discovery potential for a given detector. 
Although the couplings from the lightest Higgs boson are singlet-like many final states show a compatible cross section times branching ratio compared to the SM Higgs boson because of the large phase-space for a light Higgs boson.  
Assuming similar efficiencies and backgrounds and a discovery potential scaling with the luminosity L as $\sqrt{L}$ the red (orange) areas in Fig. \ref{f3-0} are of interest to look at for the expected integrated luminosity of 300 (3000) fb$^{-1}$. 
The whole mass range of the lightest Higgs boson is accessible with the tau final states. However, the efficiency, especially the trigger efficiency, has to be investigated for the benchmark points.
The gamma final states are also of interest to investigate but they have a larger background for lower masses.  
The final states including Z,W bosons can only be used for the high mass region above 80 GeV. Although the decay into the lightest pseudo-scalar Higgs bosons and neutralinos can have large values for the cross section times branching ratio, this decay is only possible if kinematically allowed. 
A discovery of the singlet-like Higgs boson would strongly hint towards a singlino-like dark matter candidate, which is compatible with all direct dark matter searches \cite{Beskidt:2017xsd}.

\section*{Acknowledgements}
Support from the Heisenberg-Landau program and the Deutsche Forschungsgemeinschaft (DFG, Grant BO 1604/3-1)  is warmly acknowledged.


\bibliographystyle{lucas_unsrt}
\bibliography{light-higgs-br_v12}

%

\newpage
\appendix{\bf{Appendix}}

\vspace{1cm}

We select benchmark points (BMP) from the Tables 1 and 2 from the main paper, which give the range of branching ratios. 
The benchmark points in Tables \ref{t41} and \ref{t42} have been selected to have a maximal cross section times branching ratio for the corresponding decay mode. The cross sections, branching ratios and masses for some SUSY particles have been indicated. The corresponding input parameters for NMSSMTools are given in Table \ref{t43}.

\begin{table}
[!b]
\scriptsize
\centering
\caption{Summary of benchmark points (BMP) for the two light Higgs mass bins below 40 GeV and the possible discovery channels. Values for the branching ratio and cross section times branching ratio below 0.01 are set to zero.  \label{t41}}
\begin{tabular}{l|c|c||c|c|c|c}
	\hline\noalign{\smallskip}
	& \multicolumn{2}{c||}{0-20 GeV} & \multicolumn{4}{c}{20-40 GeV} \\  
	\noalign{\smallskip}\hline\noalign{\smallskip}
	 					& BMP 1 	& BMP 2 & BMP 3 	& BMP 4 & BMP 5 	& BMP 6  \\
	\noalign{\smallskip}\hline\noalign{\smallskip}
	decay mode & $\tau\tau$ & $\gamma\gamma$   &  $\tau\tau$ & $\gamma\gamma$ & $A_1 A_1$ & $\tilde{\chi}_1^0 \tilde{\chi}_1^0$ \\
	\noalign{\smallskip}\hline\noalign{\smallskip}
 $\sigma_{ggf} \times BR $ in fb 		&  12103.9	& 1.8	& 7256.2 	&  2.4	& 4232.4 &  4639.1\\
 $\sigma_{bbH} \times BR $ in fb 		&  121.2	& 0.0 	& 313.1 	&  0.0	& 376.5	&  305.4 \\
	\noalign{\smallskip}\hline\noalign{\smallskip}
$BR(H_1 \rightarrow g g)$ in \% 		&  1.03	  	&  7.55 &    0.62	& 19.73	&  0.36  & 0.36\\
$BR(H_1 \rightarrow \tau \tau)$ in \% 		&   10.16 	&  0.47 &  7.67   	&  0.27	&  5.41	& 5.37 \\
$BR(H_1 \rightarrow c \bar{c})$ in \% 		&  0.12 	&  86.38&  0.17 	& 76.45	&  0.20 & 0.16  \\
$BR(H_1 \rightarrow b \bar{b})$ in \% 		&  88.65	&  5.42	&   91.51 	&  3.01	& 63.63 & 63.56  \\
$BR(H_1 \rightarrow \gamma \gamma)$ in \% 	&  0.00	   	& 0.18	&  0.00		& 0.54	& 0.00 & 0.00 \\
$BR(H_1 \rightarrow A_1 A_1)$ in \% 		& - 		& - 	&  - 		& - 	&  26.14& -  \\
$BR(H_1 \rightarrow \tilde{\chi}_1^0 \tilde{\chi}_1^0)$ in \% 
						& - 		& - 	&  - 		& - 	&  4.24  & 30.54   \\
	\noalign{\smallskip}\hline\noalign{\smallskip}
$m_{H1}$ in GeV 				&  15		& 20   	&  26		&  35	&  34		& 32 \\
$m_{H2}$ in GeV 				&  125		&  124 	&  122		&  124	& 122 		& 123 \\
$m_{H3}$ in GeV 				&  800		& 1351 	 &  676		& 1550 	 & 1405 	&  1951\\
$m_{A1}$ in GeV 				& 500		& 200  	&   100		&  200	&  13  		&  21\\
$m_{A2}$ in GeV 				& 798  		& 1351 	 & 675 		&  1550	& 1404 		& 1951 \\
$m_{H^\pm}$ in GeV 				&  794		& 1354 	 & 6722		& 1552	 &  1405	& 1951 \\
$m_{\tilde{t}_1}$ in GeV 			&  1126		& 943  	&  761 		& 885  	& 851 		& 820 \\
$m_{\tilde{t}_2}$ in GeV 			& 1791	 	& 1645 	&  1692	 	& 1611 	&  1722	 	& 1721 \\
$m_{\tilde{q}_L}$ in GeV 			& 2212	 	&  2222	&  2215	 	&  2224	&  2216	 	& 2220 \\
$m_{\tilde{g}}$ in GeV 				& 2240 		& 2237	 &  2241 	& 2237	 & 2240	 	& 2240 \\
$m_{\tilde{\chi}_1^0}$ in GeV 			& 250 		& 98 	&  50 		&  98	&   16 		& 10 \\
$m_{\tilde{\chi}_2^0}$ in GeV 			&  321		& 111 	&  163 		&  111	&  199		&  207\\
$m_{\tilde{\chi}_3^0}$ in GeV 			&  333		& 120 	&  181 		&  124	&  216		& 223 \\
$m_{\tilde{\chi}_4^0}$ in GeV 			&  441		& 432 	&  434 		& 432 	&  435		& 435 \\
$m_{\tilde{\chi}_5^0}$ in GeV 			& 829	 	& 825 	& 825  	 	&  826 	&  826		& 827 \\
$m_{\tilde{\chi}_1^\pm}$ in GeV 		& 295 		& 104	& 154  		& 105  	&  200		& 208    \\
$m_{\tilde{\chi}_2^\pm}$ in GeV 		& 829 		& 825 	&  8252		& 826 	&  826		& 827  \\
	\noalign{\smallskip}\hline
\end{tabular}
\end{table}

\begin{table}
[!t]
\scriptsize
\tiny
\centering
\caption{As in Table \ref{t41} but for the mass range $40$ GeV $< m_{H1} < 120$ GeV. \label{t42}}
\begin{tabular}{l|c|c|c|c||c|c|c|c}
	\hline\noalign{\smallskip}
	& \multicolumn{4}{c||}{40-90 GeV} & \multicolumn{4}{c}{90-120 GeV} \\ 
	\noalign{\smallskip}\hline\noalign{\smallskip}
	 					& BMP 7 	& BMP 8 & BMP 9 	& BMP 10 & BMP 11 & BMP 12 & BMP 13 & BMP 14 \\
	\noalign{\smallskip}\hline\noalign{\smallskip}
	decay mode & $\tau\tau$ & $\gamma\gamma$ & $A_1 A_1$ & $\tilde{\chi}_1^0 \tilde{\chi}_1^0$ & $\tau\tau$ & $\gamma\gamma$ & $A_1 A_1$ & $WW/ZZ/Z\gamma$\\
	\noalign{\smallskip}\hline\noalign{\smallskip}
	$\sigma_{ggf} \times BR$ in fb 		& 1137.0	&  47.7	&  10755.1	& 732.9	& 173.8 & 46.0  & 646.2 &  122.6/12.8/1.0\\
	$\sigma_{bbH} \times BR$ in fb 		& 202.4		&  0.0	&  1194.6	& 133.0 & 115.4	& 0.0 	& 3.9 	&  1.5/0.1/0.0\\
	\noalign{\smallskip}\hline\noalign{\smallskip}
$BR(H_1 \rightarrow g g)$ in \% 		& 0.38 		& 39.86  &  0.09	& 0.11	& 0.28 & 47.43 	& 0.49 	& 6.43 \\
$BR(H_1 \rightarrow \tau \tau)$ in \% 		& 8.00   	& 1.05 	& 1.54 		&  2.29	& 9.39 	& 0.62  & 0.29   & 6.63  \\
$BR(H_1 \rightarrow c \bar{c})$ in \% 		& 0.09		& 43.14	&  0.13 	& 0.10 	& 0.05 	& 29.98	& 0.32  & 3.29 \\
$BR(H_1 \rightarrow b \bar{b})$ in \% 		& 91.48  	& 14.52  & 17.08  	&  22.76&90.24 & 11.09  & 2.80 &   61.80 \\  
$BR(H_1 \rightarrow \gamma \gamma)$ in \% 	& 0.00 		& 1.43 	&  0.00		&  0.00	& 0.00	& 2.06 	& 0.02  & 0.27   \\
$BR(H_1 \rightarrow A_1 A_1)$ in \% 		& - 		& - 	&  81.15 	&  48.93& - 	& - 	& 95.95  & - \\
$BR(H_1 \rightarrow WW)$ in \% 			& - 		& - 	& - 		& 0.00 	& 0.00	& 8.57	& 0.13 &  19.36 \\ 
$BR(H_1 \rightarrow Z Z)$ in \% 		& - 		& - 	& - 		& - 	&  0.00	& 0.14  & 0.00  &  2.02  \\
$BR(H_1 \rightarrow Z \gamma)$ in \% 		& - 		& - 	& - 		& - 	&  0.00 & 0.09  & 0.00  &  0.16   \\
$BR(H_1 \rightarrow \tilde{\chi}_1^0 \tilde{\chi}_1^0)$ in \%
						& - 		& - 	& - 		&  25.80& - 	& - 	& - 	& - \\
	\noalign{\smallskip}\hline\noalign{\smallskip}
$m_{H1}$ in GeV 				& 40 		& 72 	&  50 		& 83  	& 95 	& 102 	& 104 	& 123 \\
$m_{H2}$ in GeV 				& 124		& 125 	&  123		&  125	&  125	& 124  	& 124	& 124 \\
$m_{H3}$ in GeV 				&  1004		& 1801 	&  1012		& 1878 	& 401	 & 1747	&  1651	& 2000 \\
$m_{A1}$ in GeV 				& 100 		& 400 	&  7 		& 16	& 75 	& 401 	& 25 	& 200 \\
$m_{A2}$ in GeV 				&  1004		& 1801 	&  1012		&  1878	& 394 	& 1747 	& 1651 	& 2000 \\ 
$m_{H^\pm}$ in GeV 				&  1000		& 1803 	& 1012	 	&  1874	& 380 	& 1749 	& 1653 	& 2002 \\
$m_{\tilde{t}_1}$ in GeV 			&  773		&  975	&  790		& 999 	& 1289 	& 931  	&  877	&  913\\
$m_{\tilde{t}_2}$ in GeV 			& 1697	 	& 1671 	& 1703 		& 1771 	& 1845	& 1690 	& 1622 	& 1668 \\
$m_{\tilde{q}_L}$ in GeV 			& 2215	 	& 2224 	&  2215		& 2217 	& 380	 & 2222	&  2224	& 2225  \\
$m_{\tilde{g}}$ in GeV 				&  2241		& 2238	&  2241		&  2239	& 2241 	& 2239 	 & 2237	& 2238 \\
$m_{\tilde{\chi}_1^0}$ in GeV 			&  48		&  98 	& 44		& 37	& 62 	& 98 	&  98 	& 98 \\
$m_{\tilde{\chi}_2^0}$ in GeV 			&  201		& 112 	&  200		& 274 	&  158	& 111 	& 108 	& 111 \\
$m_{\tilde{\chi}_3^0}$ in GeV 			&  219		& 250 	& 219 		&  295	& 175  	& 261 	& 111 	& 174  \\
$m_{\tilde{\chi}_4^0}$ in GeV 			&  435		& 432 	& 435 		&  437	& 432 	& 432 	& 432  	& 432 \\
$m_{\tilde{\chi}_5^0}$ in GeV 			& 827 	  	& 826 	&  826 		& 827 	& 822 	& 825 	& 826 	&  826 \\
$m_{\tilde{\chi}_1^\pm}$ in GeV 		&  189		& 105 	&   201		& 268 	& 108 	& 104 	&  105	& 104 \\ 
$m_{\tilde{\chi}_2^\pm}$ in GeV 		&  827 		& 826  	&  826		& 827 	& 822 	& 825 	& 826 	& 826 \\
	\noalign{\smallskip}\hline
\end{tabular}
\end{table}

\begin{table}
[!t]
\scriptsize
\centering
\caption{NMSSM input parameters for BMP 1-14, listed in Tables \ref{t41} and \ref{t42}, for NMSSMTools version 5.2.0. \label{t43}}
\begin{tabular}{l|c|c|c|c|c|c|c}
	\hline\noalign{\smallskip}
	 & $\tan\beta$ & $A_0$ & $A_\kappa$ & $A_\lambda$ & $\lambda$ & $\kappa$ & $\mu_{eff}$  \\
	\noalign{\smallskip}\hline\noalign{\smallskip}
BMP 1	&  2.52 &  -1192.15	&  -164.93	&  332.61 &  6.04$\cdot 10^{-1}$ 	&  2.82$\cdot 10^{-1}$ 	& 299.50\\
BMP 2 	& 26.83  &  -2532.10	&  -227.59	& -505.96 &  2.19$\cdot 10^{-2}$  &  1.24$\cdot 10^{-2}$	& 103.48\\
BMP 3   & 4.01 &  -2685.78 	&  -62.95 	& -296.47 &  4.39$\cdot 10^{-1}$	&  6.41$\cdot 10^{-2}$	&  156.32   \\
BMP 4 	& 28.40  & -2638.61 	&  -220.86	&  -430.99&   2.05$\cdot 10^{-2}$	&  1.19$\cdot 10^{-2}$ 	& 103.80  \\
BMP 5 	&  6.63	& -2556.74	& 222.89  	&  707.12 &  3.00$\cdot 10^{-1}$ 	&  	1.01$\cdot 10^{-2}$ &  200.92 \\
BMP 6 	&   9.05 & -2550.05	&  328.08 	&  1352.94 &  3.00$\cdot 10^{-1}$ 	& 6.07$\cdot 10^{-3}$  & 208.14\\
BMP 7 	&  5.03	& -2588.24 	&   367.99	& 353.02	&   5.20$\cdot 10^{-1}$	& 6.05$\cdot 10^{-2}$	& 190.51  \\
BMP 8 	&  25.68&  -2408.21 	&  -444.13  	& 75.25	&  9.94$\cdot 10^{-2}$	&  1.16$\cdot 10^{-1}$ 	& 103.56 \\
BMP 9 	&  4.54	&  -2649.07	&  169.93	&  197.89 	&  3.36$\cdot 10^{-1}$ 	&  3.33$\cdot 10^{-2}$ 	& 202.86  \\
BMP 10 	&  6.71 &  -1635.23	&  2185.58	&  2658.79	&   5.62$\cdot 10^{-1}$ 	&  3.44$\cdot 10^{-2}$	&  268.28  \\
BMP 11 	& 3.50 &  -723.07  	& 584.82  	&  275.78 	&  6.68$\cdot 10^{-1}$ 	& 2.74$\cdot 10^{-1}$  	& 110.34  \\
BMP 12 	& 20.98	&  -2494.09  	& -422.59   	& 310.11	&  9.29$\cdot 10^{-2}$ 	& 1.14$\cdot 10^{-1}$ 	& 103.53 \\
BMP 13 	& 27.00  & -2639.49 	& -3.76 	&  -231.51	&  2.36$\cdot 10^{-2}$	& 1.20$\cdot 10^{-2}$ 	& 103.78 \\
BMP 14 	& 23.68  & -2518.36	& -157.16  	&   524.71	&  4.93$\cdot 10^{-3}$ 	& 4.04$\cdot 10^{-3}$	& 103.30  \\
	\noalign{\smallskip}\hline
\end{tabular}
\end{table}

\end{document}